\documentclass[conference]{IEEEtran}
\IEEEoverridecommandlockouts
\usepackage{cite}
\usepackage{amsmath,amssymb,amsfonts}
\usepackage{algorithmic}
\usepackage{graphicx}
\usepackage{textcomp}
\usepackage{xcolor}
\usepackage{bbm}
\usepackage{enumitem}

\usepackage{pifont}
\usepackage{mathtools}

\usepackage{microtype}
\usepackage{hyperref}
\usepackage{url}
\usepackage{graphicx}
\usepackage{multirow}
\usepackage{algorithm}
\usepackage{subcaption}
\usepackage{colortbl}
\usepackage{xcolor}
\usepackage{wrapfig}
\usepackage{comment}
\newcommand{\norm}[1]{\left\lVert#1\right\rVert}
\def\BibTeX{{\rm B\kern-.05em{\sc i\kern-.025em b}\kern-.08em
    T\kern-.1667em\lower.7ex\hbox{E}\kern-.125emX}}
\begin{document}

\title{Data Leakage via Access Patterns of Sparse Features in Deep Learning-based Recommendation Systems}

\author{\IEEEauthorblockN{Hanieh Hashemi\footnotemark *}
\IEEEauthorblockA{
\textit{University of Southern California}\\
hashemis@usc.edu \\
}
\and
\IEEEauthorblockN{Wenjie Xiong}
\IEEEauthorblockA{ 
\textit{Meta AI/Virginia Tech}\\
wenjiex@fb.com \\
}
\and
\IEEEauthorblockN{Liu Ke\footnotemark *}
\IEEEauthorblockA{
\textit{Washington University in St. Louis}\\
ke.l@wustl.edu  \\
}
\and
\IEEEauthorblockN{Kiwan Maeng\footnotemark *}
\IEEEauthorblockA{ 
\textit{Pennsylvania State University}\\
kvm6242@psu.edu \\
}
\and
\IEEEauthorblockN{Murali Annavaram}
\IEEEauthorblockA{ 
\textit{University of Southern California}\\
annavara@usc.edu \\
}
\and
\IEEEauthorblockN{G. Edward Suh}
\IEEEauthorblockA{ 
\textit{Meta AI/Cornell University}\\
edsuh@fb.com \\
}
\and
\IEEEauthorblockN{Hsien-Hsin S. Lee\footnotemark *\thanks{*This work was conducted while the authors were employed at Meta.}}
\textit{Intel}\\
\IEEEauthorblockA{linear@acm.org\\
}
}
\maketitle

\begin{abstract}
Online personalized recommendation services are generally hosted in the cloud where users query the cloud-based model to receive recommended input such as merchandise of interest or news feed.
State-of-the-art recommendation models rely on sparse and dense features to represent users' profile information and the items they interact with. Although sparse features account for $99\%$ of the total model size, there was not enough attention paid to the potential information leakage through sparse features. These sparse features are employed to track users' behavior, {\em e.g.,} their click history, object interactions, etc., potentially carrying each user's private information. Sparse features are represented as learned embedding vectors that are stored in large tables, and personalized recommendation is performed by using a specific user's sparse feature to index through the tables. Even with recently-proposed methods that hides the computation happening in the cloud, an attacker in the cloud may be able to still track the access patterns to the embedding tables. This paper explores the private information {that may be learned by tracking a recommendation model's} sparse feature access patterns. We first characterize the types of attacks that can be carried out on sparse features in recommendation models {in an untrusted cloud}, followed by a demonstration of how each of these attacks leads to extracting users' private information or tracking users by their behavior over time. 
\end{abstract}

\section{Introduction}
Deep learning based personalized recommendation models comprise around 80\% of AI inference cycles in production-scale data centers such as Meta, Alibaba, Amazon, and Google~\cite{covington2016deep, gupta2020architectural, chui2018notes, wu2019deep} and the trend continues to grow. They are also responsible for driving around 35\% of Amazon’s revenue~\cite{gupta2020architectural} and 80\% of hours streamed on Netflix~\cite{gomez2015netflix}. These models use many types of information, including user attributes, user preferences, user behavior, social interaction, contextual information, and others~\cite{erkin2010privacy} to provide personalized recommendations relevant to a given user.  

Early recommendation systems such as content filtering approaches~\cite{lang1995newsweeder} rely on users' explicit indication of their own interests and preferences to match items for each user. In contrast, modern recommendation systems embrace machine learning such as Deep Learning Recommendation Models (DLRM) which rely on both dense (continuous) features and sparse (categorical) features to track user's past behavior and learn their preferences. While such implicit knowledge generation improves user experiences, 
{the deep learning based recommendation also rely on more user data, which need to be protected for privacy.}

Due to the ever-growing concern for data privacy, governments' regulations such as HIPPA~\cite{annas2003hipaa}, European GDPR~\cite{di2022recommender}, and California Privacy Act~\cite{rothstein2019california} have been enacted, requiring service providers to protect and limit the collection of certain personal sensitive information. 
{If such a sensitive information is collected and used by the service provider, a malicious attacker in the cloud~\cite{gruschka2010attack} may be able to infer private user information. This paper investigates the potential information leakage through sparse feature accesses when recommendation systems use an untrusted cloud and the features are leaked.}

 \begin{figure}[t]
 \centering
 \includegraphics[scale = 0.131]{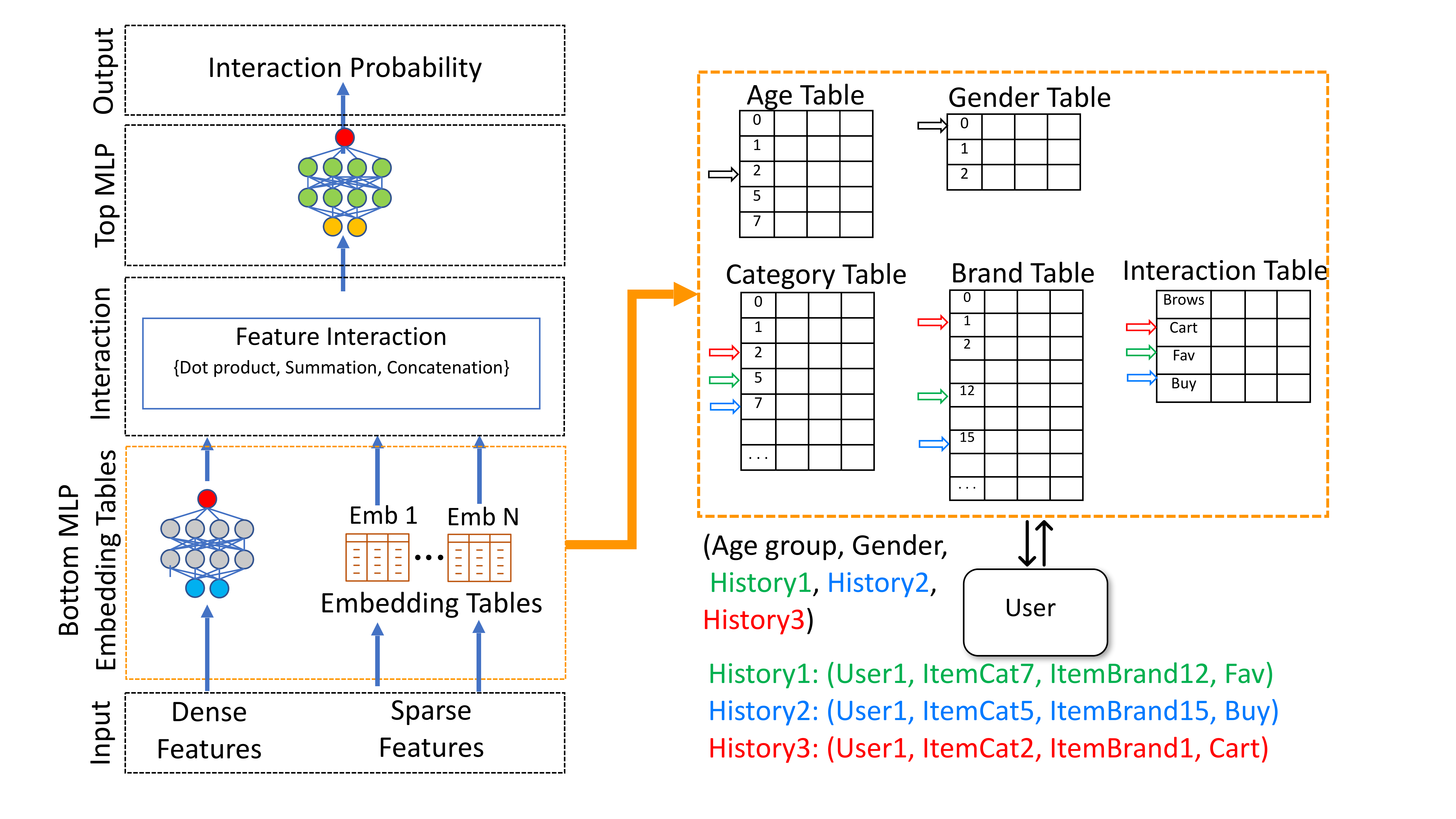}
  \vspace{-5mm}
 \caption{left: DLRM model architecture, right: An example of embedding table processing.}
 \label{fig:dlrm}
\end{figure}

There are several commercially used deep learning based recommendation models such as DLRM and TBSM~\cite{naumov2019deep, ishkhanov2020time} from Meta, Wide and Deep~\cite{cheng2016wide} from Google, and DIN and DIEN~\cite{zhou2018deep, zhou2019deep} from Alibaba. They all utilize both dense and sparse features under a similar model architecture for making personalized recommendations. To understand the privacy challenges, we provide a brief illustrative background using DLRM~\cite{naumov2019deep}, a state-of-the-art deep learning based personalized recommendation model from Meta as an example. As shown in Figure~\ref{fig:dlrm}, there are five types of sparse features in this example: age, gender, item category, item brand, and the user interaction type (such as an item purchased, an item marked as a favorite, etc.). The number of rows in each table is equal to the number of categories in each feature. The sparse features must be first converted into dense representations in the form of embedding so that semantically similar features are represented by similar embedding values. This conversion is done by the embedding tables present in recommendation models~\cite{naumov2019deep,ishkhanov2020time, cheng2016wide}. To get an accurate recommendation, some of the past interactions of the user, user gender, and age data are used as user features. The sparse features are used to generate corresponding indices into the embedding tables to perform embedding lookup and the follow-up embedding reduction and aggregation. For instance, the age information is used to index the age table, while each of the three category values from the three past interactions is used to access the category table. Note that in general there are two types of sparse features. The sparse features that remain unchanged or are modified infrequently (e.g., age or gender) are referred to as {\em static sparse features} while the features that are changed frequently such as information related to items browsed and purchased by a user are called {\em dynamic sparse features}.

Production-scale recommendation systems usually need to process and make recommendations through a very large number of items; hence they demand a large amount of memory, {\em e.g.,} tens of TB~\cite{Mudigere2021, sethi2022recshard}, to accommodate these large embedding tables. A recent study showed that $99\%$ of the model parameters belong to the embedding tables~\cite{gupta2020architectural}. 
The goal of this paper is to {study} the types of information leakages via sparse feature accesses.
While attacks on dense features have been explored in several prior research investigations in different ML domains such as computer vision~\cite{akhtar2018threat, choquette2021label, li2021membership}, speech~\cite{abdullah2021sok, feng2021attribute, schonherr2018adversarial, feng2022enhancing}, natural language processing~\cite{zhang2020adversarial, carlini2021extracting, chen2021badnl}, and graph neural networks~\cite{he2021stealing, zhang2021backdoor}, less attention was paid to the potential leakage through sparse features constituting embedding table accesses. This paper demonstrates different types of {private data leakages that can} occur {in the cloud-based} deep learning recommendation systems through sparse features. 

Some of the attacks we discuss in this paper such as the identification attack and the sensitive attribute attack were investigated in other recommendation models such as content filtering, but to the best of our knowledge this paper is the first effort analyzing these {threats} in deep learning based recommendation models. Furthermore, the re-identification attack and the frequency-based attack are only explored in other domains such as databases. This paper {investigates} these attacks in the recommendation system domain. Also, we show that even secret hash functions cannot {completely} solve the privacy issue. The main contributions of this work are:
\begin{itemize}[noitemsep, leftmargin=*]
    \item {\bf Identification attack.}
    We demonstrate that the collection of seemingly innocuous, isolated features from a user's perspective can potentially become a highly narrow representation of a small group of individuals, thereby leading to an identification attack. 
    \item {\bf Sensitive attribute attack.}  We show how users' past behaviors are highly correlated to some of their sensitive sparse features (such as age and gender). Even if the users are not concerned about leaking their past history, they may not want to leak information about their sensitive sparse features such as gender. However, even without explicitly sharing the sensitive features, it is possible for an adversary to infer these features by just observing the user's past behavior.
    \item {\bf Re-identification attack.} We also demonstrate that the history of recent purchases when disclosed to an adversary in the cloud can lead to re-identification of users who send queries to the recommendation model. For instance, the most recent purchases of a user provide a {fairly good} fingerprint of the user. Such a fingerprint can be used by an attacker to track and re-identify a user over time and across multiple interaction sessions.
    \item {\bf Frequency-based attack.} Industrial implementations of embedding tables use hashed representation to map multiple feature values to a single embedding entry.  We demonstrate that such hash operations do not noticeably reduce the amount of information. Our study shows that an adversary can rebuild the raw sparse features with high accuracy by observing the frequency of accesses for each embedding table entry. 
    \item {\bf {Private hash leakage}} We take a step further and replace the current hash functions that are used for efficiency purposes with a {secret} hash that maps pre-hash entries to a random post-hash entry. We also assume that the attacker does not have any information about the hash function and tries to reverse-engineer it with frequency analysis. We designed an attack using machine learning optimizations to show even in this worst-case scenario attackers can extract the raw feature values with high accuracy. 
\end{itemize}
The rest of the paper is organized as follows. In Section~\ref{sec:background}, we discuss related work, threat model, and datasets. Section~\ref{sec:ident} and Section~\ref{sec:sensitive} discuss identification and sensitive attribute attacks. The re-identification attack is elaborated in Section~\ref{sec:reIdent}. In Section~\ref{sec:pipeline}, we show the data pipeline in production-scale models and formulate the information leakage through sparse features. 
In Section~\ref{sec:design}, we illustrate how the raw value of a sparse feature can be recovered by the frequency-based attack. In Section~\ref{sec:privateHash}, we designed an advanced attack to demonstrate that an attacker can break a private hash using the frequency of accesses. We discuss the implications to private recommendation systems and the related works in Section~\ref{sec:future} and~\ref{sec:imp}. In Section~\ref{sec:con}, we draw the conclusion. 

\section{Background and Threat Model}
\label{sec:background}
\subsection{Deep Learning Based Recommendation Systems}
 One of the earliest recommendation systems is the content filtering model~\cite{lang1995newsweeder} in which experts classify items and assign meta-data to them. Users also mark their interests, and recommendations are done based on their preferences. 
 In collaborative filtering~\cite{walker1967estimation}, recommendations are based on past behaviors. The assumption is that similar users like similar items or users like items similar to their highly-rated items. In neighboring methods~\cite{goldberg1992using}, the algorithm suggests recommendations by grouping users and products together. 
 
 Recently, predictive analytics models have attracted more attention. These models include simple models such as linear and logistic regression~\cite{frolov2017tensor} to deep learning based recommendation systems~\cite{wu2017recurrent, zhang2019deep,hansen2020contextual, okura2017embedding}. Netflix showed DNN-based recommendation systems can shine when heterogeneous types of data are used~\cite{steck2021deep}. Recently the use of deep learning based recommendation models that can use both dense and sparse features are highlighted by DLRM and TBSM~\cite{naumov2019deep, ishkhanov2020time} from Meta, Wide and Deep~\cite{cheng2016wide} from Google, and DIN and DIEN~\cite{zhou2018deep, zhou2019deep} from Alibaba. Typically, the dense features represent some user characteristics with floating point values. Examples of dense features include historical statistics like link click ratio, page connect latency, or timestamp of the past user behaviors, while the sparse features represent categorical information such as interests and user past behavior information (user/item interactions). As shown in Figure~\ref{fig:dlrm} in DLRM, continuous features are learned using Multi-Layer Perceptrons (MLPs) while the sparse features are learned through embedding tables that map them to dense representations~\cite{naumov2019deep}. It is clear from the history of recommendation systems that as models progressed over time they increasingly rely on implicitly gained user information to provide accurate predictions, rather than relying on explicit user preferences. As such, it is necessary for users of these systems to understand how adversarial behavior in the cloud can compromise their private information, some of which they may not even know their existence.  

\subsection{Threat Model}
\begin{figure}[t]
 \centering
 \includegraphics[height=1.8in,width=3in]{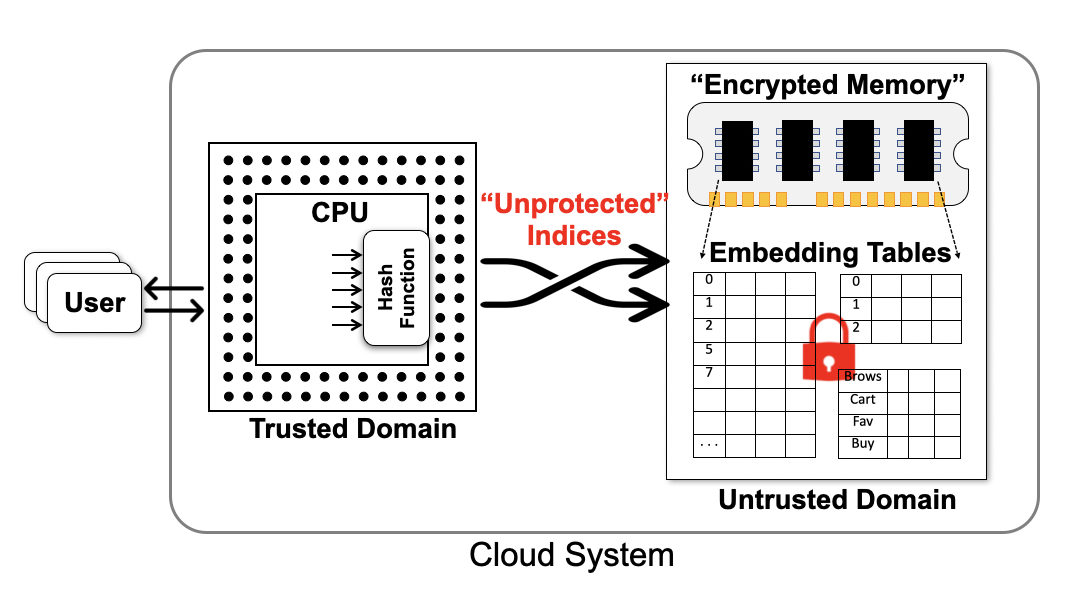}
  \vspace{-2mm}
 \caption{Users share their data with cloud. While computations are protected, access patterns are vulnerable to attacks.}
 \label{fig:threat}
\end{figure}

As shown in Figure~\ref{fig:threat}, users are data owners who consent to share their private data to receive recommendations from a cloud-based model. However, by sharing data to the cloud, undesired/unauthorized usage of data may still take place when an intruder or an untrusted employee observes an inference query request. We assume that the computation and the query data are protected, and do not consider direct information leakage from them. The reason for this assumption is that protecting dense features {can be done efficiently through} various techniques (in Section~\ref{sec:imp}).  We also assume all the communication channels are encrypted ({\em e.g.}, using TLS).
However, {today's secure computing techniques do not prevent the leakage of sparse features through embedded table access patterns unless expensive cryptographic protection with significant overhead is added}. 
In that sense, we assume that an attacker can obtain the indices that are used to access embedding tables in a recommendation model. {Note that our threat model considers an adversary who is far stronger than what is typically assumed in today's cloud threat model so that we can better understand potential threats when cloud providers cannot be fully trusted. Most attacks in this paper will not apply to the cases when a cloud infrastructure can be trusted to provide strong isolation and prevent an adversary from observing memory access patterns to embedding tables.}

\subsection{Data sets}
\label{sec:dataset}
For studying the attacks in the following sections, we use multiple open source datasets such as Taobao Ads Display, Kaggle Ads Display, and Criteo Display. In this section, we briefly explain the content of these datasets, and in the each of following sections, we explain more details about the dataset characteristics. 

\textbf{Taobao Ads Display~\cite{Taobao}}: This dataset contains 
user static features that includes $1,140,000$ users and 10 static features per user including their user IDs. There are also other features representing a user's profile, {\em e.g.}, age, gender, occupation level, living city, education level, etc. Another file contains user behavior data that includes seven hundred million records of users past behaviors. It contains shopping behavior over 22 days. Each row of this file indicates an interaction between a user (represented by user ID) and an item (represented by item brand ID and category ID). The type of interaction (buy, brows, fav, cart) and the time stamp of the interactions.

\textbf{Kaggle Ads Display~\cite{Kaggle}}: CriteoLabs shared a week’s worth of data for you to develop models predicting ads' click-through rates (CTR). This dataset contains three data files including training and test files. The training file consists of a portion of Criteo's traffic over a period of 7 days. Each row corresponds to a display ad served by Criteo. Positive (clicked) and negative (non-clicked) examples have both been subsampled at different rates to reduce the dataset size. Each row contains 13 dense features and 26 sparse features that form embedding table accesses. The semantics of these features is not released. 
The test set is computed in the same way as the training set but for events on the day following the training period. 

\textbf{Criteo Ads Display~\cite{Criteo}}: This dataset is similar to Kaggle. But it is a much larger dataset containing 24 data files collected over 24 days with a different subsampling ratio. 

For the identification attack, sensitive attribute attack, re-identification attack, and OMP-based frequency attack our analysis requires user IDs, static profile features, or user past behaviors in the same dataset. Hence, for these attacks, we used the Taobao dataset, which is the only public dataset containing all these features. For the frequency-based attack, we need less information to implement the attacks. Thus all the datasets meet the requirement and we evaluate all of them in the hash information leakage study and the frequency based attack.

\section{Identification Attack with Static User Features} 
\label{sec:ident}
\begin{figure}[t]
 \centering
 \includegraphics[height=1.8 in]{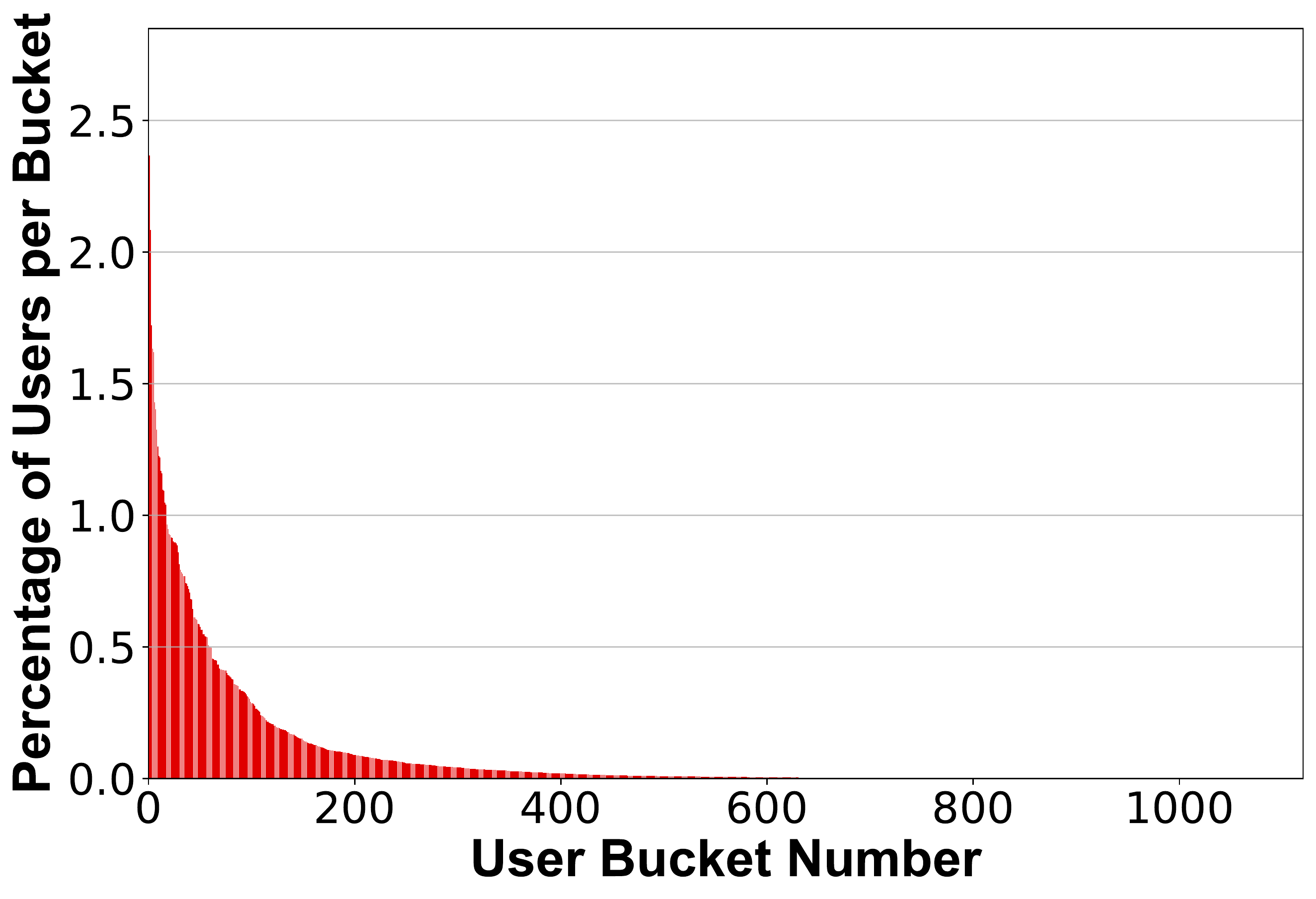}
  \vspace{-2mm}
 \caption{Percentage of the users belong to each user bucket.}
 \label{fig:UserGroup}
\end{figure}
\begin{table*}[t]
  \vspace{-2mm}
\caption{The number of users with anonymity level below K in the identification attacks (out of $1.14$ million users).}
\label{tab:anonym}
\centering
\resizebox{2\columnwidth}{!}{%
\begin{tabular}{c ccccccccc}
\hline
{\bf 1-anonymity} & {\bf 2-anonymity} & {\bf 3-anonymity}   & {\bf 4-anonymity} & {\bf 5-anonymity} &{\bf 6-anonymity} &{\bf 7-anonymity}&{\bf 8-anonymity}& {\bf 9-anonymity} & {\bf 10-anonymity}\\ \hline
$56$&$154$&$256$&$380$& $480$ &$606$ &$739$&$867$&$984$&$1104$\\ \hline

\centering
\end{tabular}
}
\end{table*}
Recall that a single user's inference request contains a series of sparse features, each of which in isolation has limited user information. However, multiple sparse features together can form a distinctive fingerprint for personal identification. 

User profile attributes (e.g. gender, city, education, etc) are usually static, in other words, they do not change or the frequency of the change is extremely low. We categorize this type of features into two subcategories---identifiable features and unidentifiable features. For example, email address, phone number, zip code, etc. can reveal the \textit{identity} of the user~\cite{lam2006you}. However, because of strict regulations in many domains, most of the recommendation systems do not collect and use such sensitive information. For example, HIPPA requires organizations to remove these identifiers from datasets by a standard de-identification process~\cite{chevrier2019use}. The question is if \textit{unidentifiable} features such as age, gender, education level, and shopping history can provide sufficient information to identify a user. \\
\noindent \textbf{Evaluation Setup:} To answer this question, we analyzed an open-source dataset released by Alibaba. Taobao Ads Display dataset is a dataset of click rate prediction about display Ads, which are displayed on the website of Taobao. This dataset contains static user features including user ID (1.14M), micro group ID (97), group ID (13), gender (2), age group (7), consumption grade/plevel (4), shopping depth (3), occupation/is college student (2), city level (5). The numbers in the parenthesis are the total number of distinct values allowed for each of these features. These categories are generally self-explanatory, but the precise definition of each of these categories is listed in the dataset.  

\noindent \textbf {Attack Method} Given this collection of static features, the only directly identifying feature associated with a single user is the user ID. After removing the user ID, the collection of all other features provides $2.1$ million possible choices. Hence, after removing the user ID, a user may mistakenly think that he or she is anonymous, and revealing any of the other features to the attacker on its own will not reveal the identity of the user. 
However, based on the user profile information from more than 1~million users released in the same dataset, it is observed that in the real world only $1120$ combinations of these static feature values are possible. We refer to this $1120$ as \textit{user buckets}. We plotted the histogram of users in these $1120$ buckets as shown in Figure~\ref{fig:UserGroup}. The x-axis in the figure indicates the bucket number ($[1-1120]$) and the y-axis shows the percentage of users per bucket. This histogram is quite illuminating in how the user distributions follow a long tail pattern. In particular, there are only a few users in buckets $600$ to $1120$. In fact, there are only $989$ users on average across all these buckets, and the last $56$ buckets have only $1$ user. Consequently, observing the entire combinations of seemingly innocuous features from each may allow an attacker to launch an \textit{identification attack} to extract the unique user ID with very high certainty. In summary, without access pattern protection an attacker can observe the combination of sparse feature values of a user and based on this combination, they can find out which user bucket this user belongs to. In buckets with only a few users, they can link the query to the user who sent it with high accuracy.  

\noindent \textbf{Evaluation Metric:} For our analysis, we used a well-known property known as {\em K-anonymity} used in information security and privacy. Based on this metric, if a user's bucket number is revealed and there are K users in the same bucket, the probability of finding the user is $\frac{1}{K}$. For instance, 1-anonymity for a user means that this is the only user having this particular set of feature values.

\noindent \textbf{Evaluation Result:} As shown in Table~\ref{tab:anonym} which is extracted from the Taobao dataset,
for $56$ of the user buckets, there is only one user with a specific combination of static features which implies that an attacker can identify these users with 1-anonymity if they can observe this combination of feature values. Also for more than $1000$ users, the anonymity level is 10 or below.

\section{Sensitive Attribute Attack by Dynamic User Features}
\label{sec:sensitive}
The prior section shows how a user can be de-anonymized with multiple feature combinations. In this section, the question is when the user removes the static features, can sensitive features leak through other non-sensitive features? For instance, a user may provide no age information or provide wrong age information as a static feature. In that case, the user may have a sense of protecting more of their private data since they have not disclosed their static features. However, in this section, we demonstrate that even when a user hides their sensitive static features, adversaries are still able extract the sensitive attributes through cross correlations with user-item interaction data. \\
\noindent \textbf{Evaluation Setup:} In this part, we use dynamic sparse features. These features include user-item interactions~\cite{zhao2019preference} in the Alibaba Ads Display dataset. This dataset contains $723,268,134$ tuples collected between $4/21/2017$ to $5/13/2017$. Each tuple includes a user ID ($1.14M$), a btag (4: browse, cart, favor, buy), a category id ($12K$), and a brand ($379K$). \\
\begin{figure}[t]
 \centering
 \includegraphics[height=1.2in,width=3.4in]{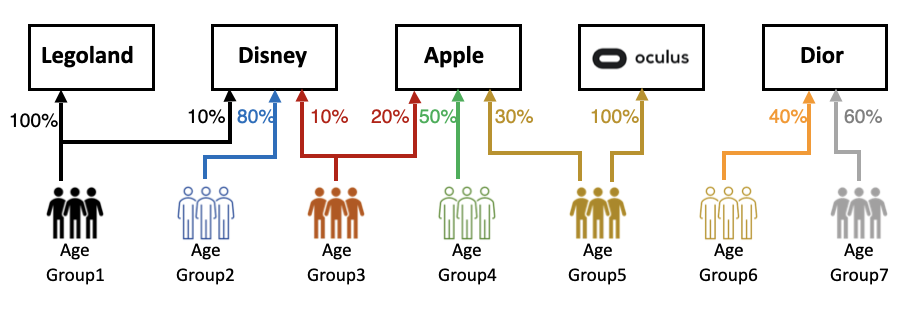}
  \vspace{-2mm}
 \caption{Different brands are popular between different customer age groups (e.g. all accesses to Legoland are from 1 age group while Apple has customers from 3 different age groups.)}
 \label{fig:Diversity}
\end{figure}
\begin{figure*}[t]
 \centering
\includegraphics[height=1.5in,width=2.2in]{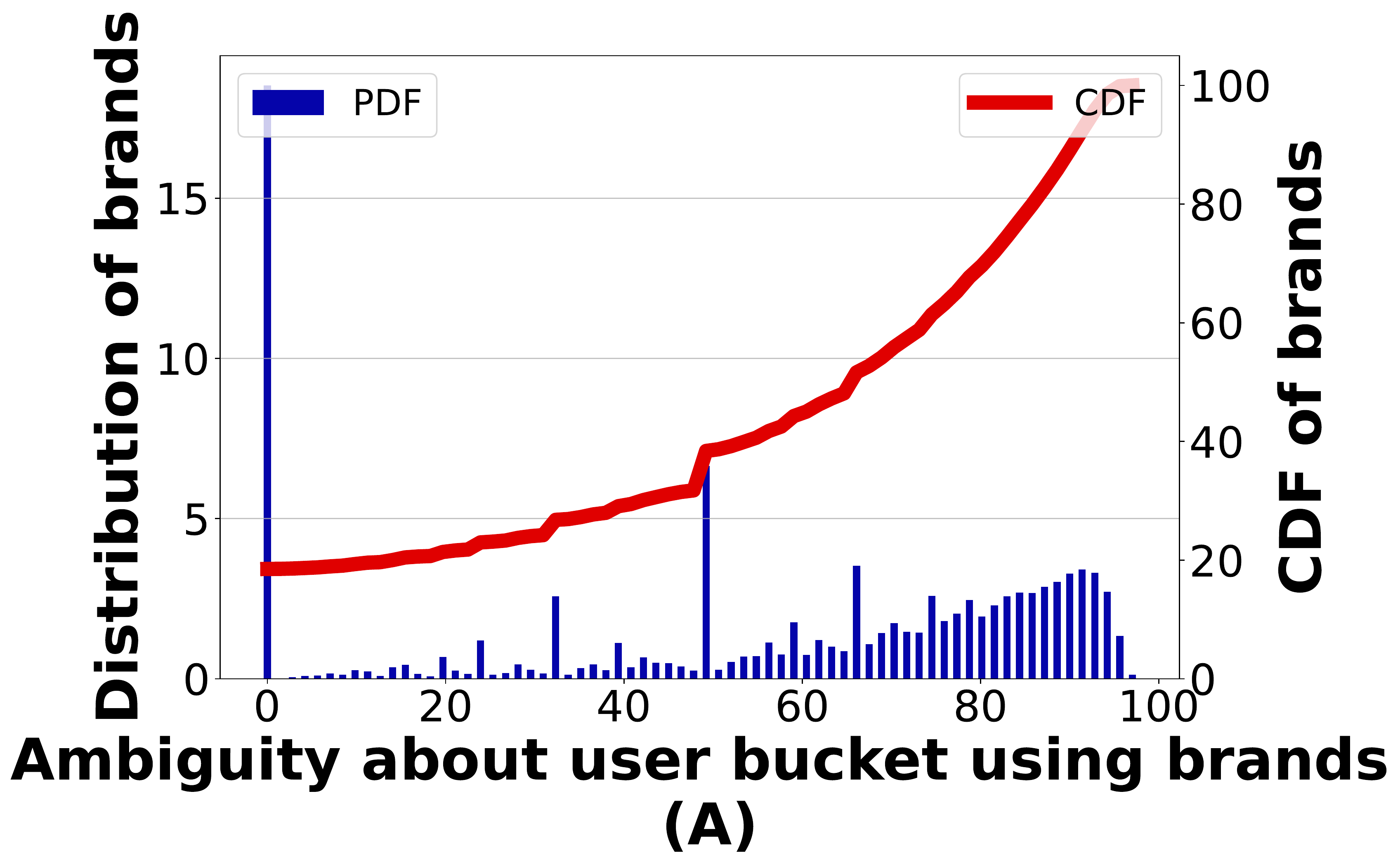}\hfill
\includegraphics[height=1.5in,width=2.2in]{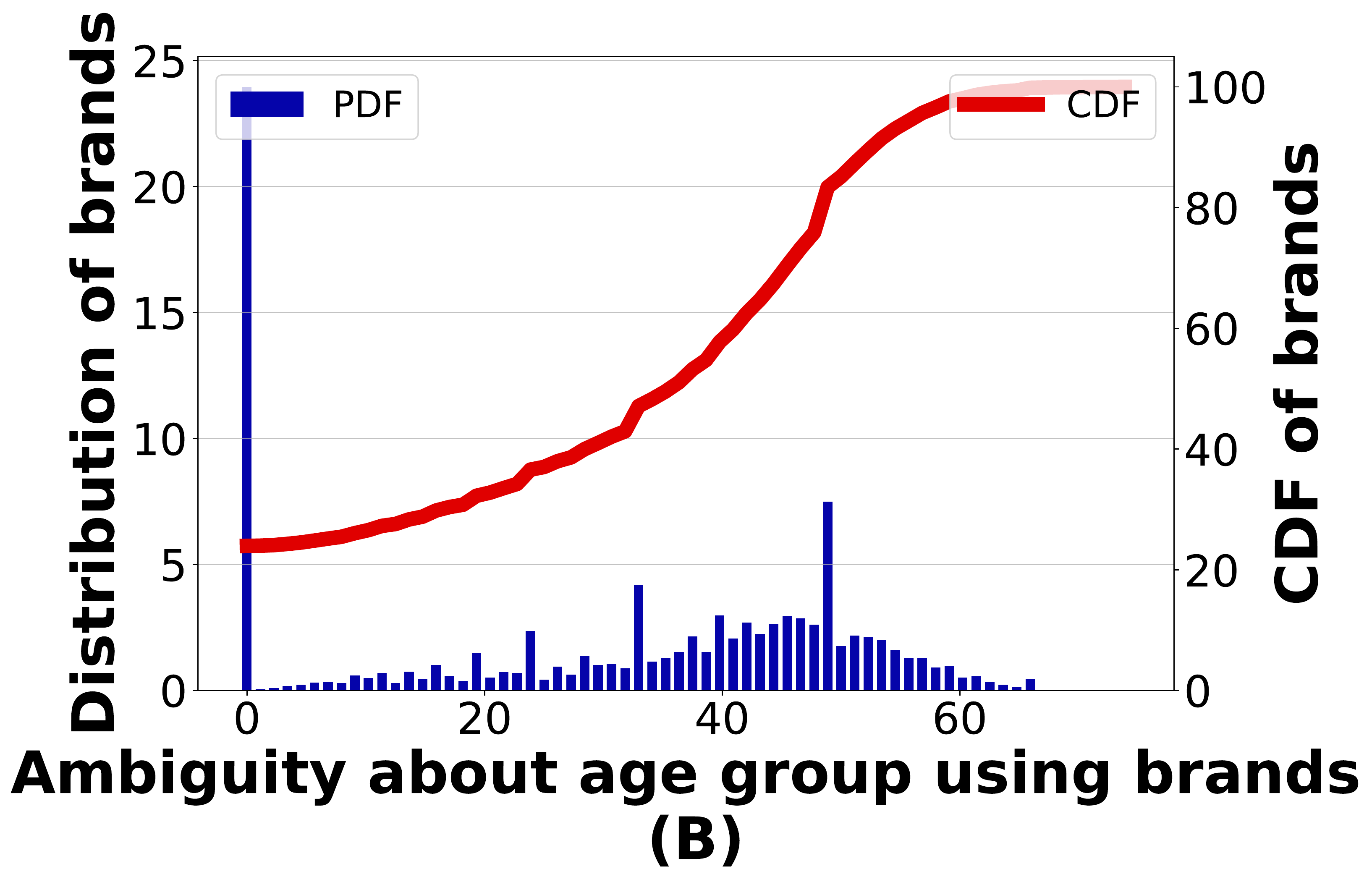}\hfill
\includegraphics[height=1.5in,width=2.2in]{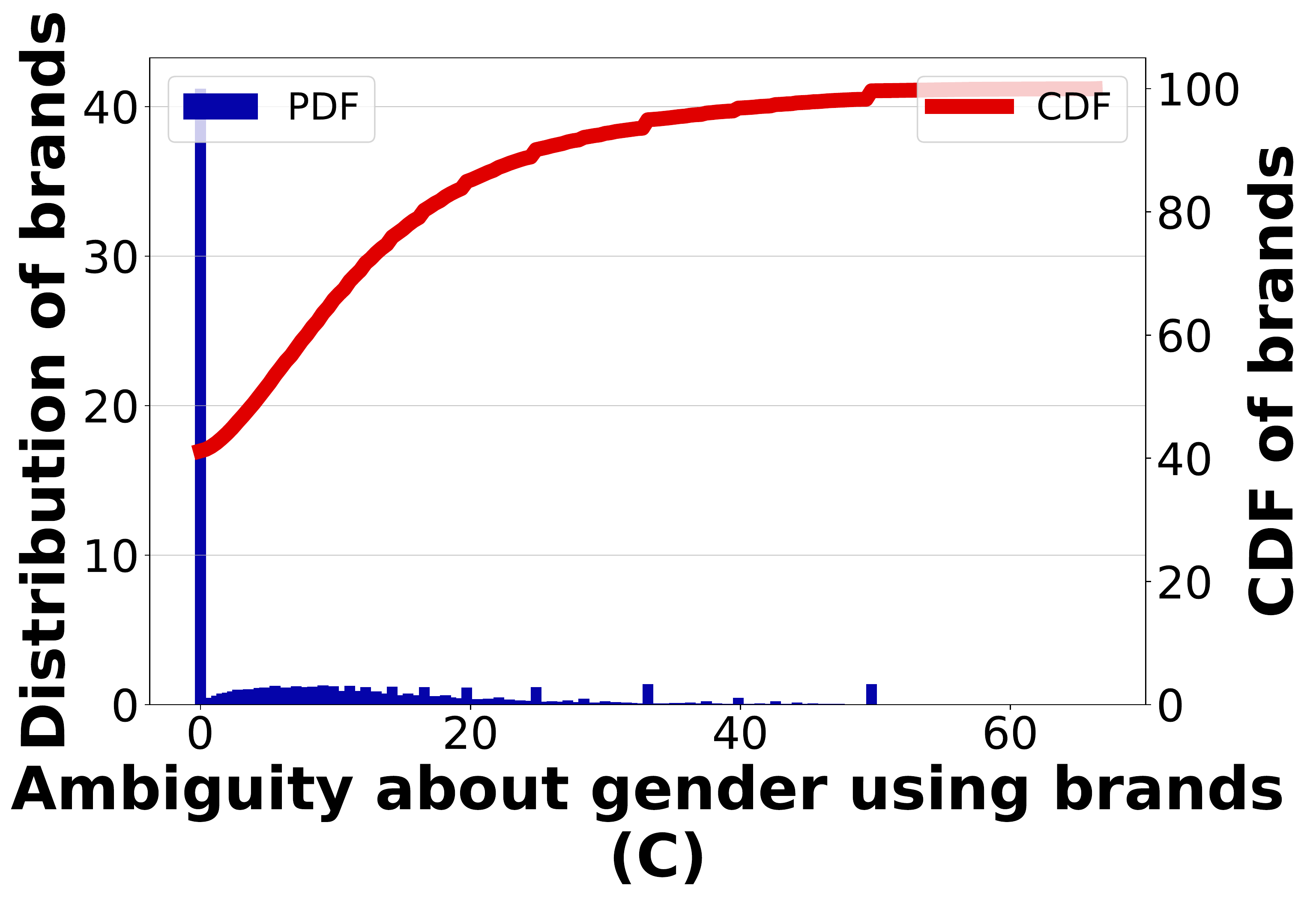}
\caption{Using the accessed brands, ambiguity about A) user buckets (defined in previous section), B) user age groups, and C) user gender groups.}
 \label{fig:brand}
\end{figure*}
\noindent \textbf{Attack Method:} Figure~\ref{fig:Diversity} depicts an example of how different brands of the items are accessed by different user age groups. The user/item interactions are depicted as graphs where each edge weight represents the fraction of the total interactions with that specific item from the corresponding age group. What is interesting to note is that in this real-world dataset, there are certain brands, where users from just a single age group interact with, in this example Legoland. In this particular example, a user who wants to protect their age group may not provide their age, but the adversary may deduce their age with a high probability if the user interacted with Legoland. While this simple illustration highlights the extremity (only one age group interacting with an item), this approach can be generalized. In General attacker in this section, uses their prior knowledge on the popularity of the items between different demographic groups. Then when they observe online queries coming from users, based on this prior information they link the query to the demographic who formed most of the accesses to that item. \\
\noindent \textbf{Evaluation Metric:} In this part, we employ a metric called \textit{ambiguity} to determine the likelihood an adversary \textit{fails} to predict a user's static sparse feature by just viewing their interactions with items.  We define ambiguity for each item $i$ as: $ambiguity_i = 100\%-max(frequency_i)$ where $frequency_i$ is the distribution vector of all accesses to brand $i$ by different user groups. Using Figure~\ref{fig:Diversity} as an example, $frequency_{apple} = [0,0,20\%,50\%,30\%,0,0]$ and as a result $ambiguity_{Apple} = 50\%$, meaning if a user has interacted with item $i$ (Apple), the attacker can predict the static feature (age group) successfully for $50\%$ of the users. With this definition, $ambiguity_i = 0$ indicates if a user has interacted with item $i$, the attacker can successfully determine the user's sparse feature.\\
\noindent \textbf{Evaluation Result:} We plotted the ambiguity metric in Figure~\ref{fig:brand}. We quantify the ambiguity of predicting a user's sparse feature, such as age and gender, by using their item (brand) interaction history alone. The x-axis of these figures shows the percentage of ambiguity where a value of 0 indicates that there is no ambiguity, and this brand is always accessed by only one user bucket. On the other hand, higher values indicate more ambiguity, and hence brands with higher values on the x-axis are popular across multiple user buckets. We plot both probability density function (PDF) and cumulative distribution function (CDF) of the ambiguity of different brands.
What is revealing in the data is that in Figure~\ref{fig:brand}(A), we observe that more than $17\%$ of brands are only accessed by 1 user bucket represented by the leftmost tall bar of PDF, meaning the attacker can determine the user bucket using those brands interactions. As shown in the CDF curve in Figure~\ref{fig:brand}(A), for $38\%$ of the brands, the attacker can predict the user bucket with a success rate of greater than $50\%$. 

We present the information of age group versus ambiguity in Figure~\ref{fig:brand}(B). We observe that $24\%$ of the brands are only accessed by 1 age group. This means for $24\%$ of brands, by observing just the brand of an accessed items, one can predict the user age correctly. We did the same analysis for the gender and observed that more than $41\%$ of the brands are only accessed by one gender (Figure~\ref{fig:brand}(C)), and for $84\%$ of the brands the attacker can predict the user gender with less than $20\%$ ambiguity. \textit{Thus, even if the user avoids including any static sparse features in their recommendation inference requests to the cloud service, an adversary can extract the information with relatively low ambiguity ({\em i.e.}, high certainty).}
\section{Re-Identification Attack}
\label{sec:reIdent}
In re-identification attack, an attacker tracks users across multiple inference sessions. The goal of an attacker is to identify the same user over time by just observing their interaction history. Studies have shown the majority of the users prefer not to be tracked even anonymously~\cite{teltzrow2004impacts}. The attacker in this scenario does not try to find the identity of the user, instead, they want to track whether the same user is interacting with the recommendation system over time. In this section, we first study if the history of the purchases of a user can be used as a tracking identifier for the user. Hence, we analyze if the history of the purchases which is part of user features shared with a recommendation system is unique for each user. Second, we study if an attacker can re-identify the same user who sent queries over time by only tracking the history of their purchases, with no access to the static sparse features. We analyze the accuracy of the attacker for re-identifying the users.
\noindent \textbf{Evaluation Setup:} To study if the history of purchases for a user can be used as an identifier for the user, we extracted interaction information from the user behavior of the Taobao dataset. In the dataset, there are more than $723$ million user-item interactions. Within that interaction list, we separated about $9$ million interactions where a user purchased the item they interacted with. We then pre-processed and sorted all purchases for each user separately based on the timestamps. We then formatted that data in a time series data structure as follows for all the users and all the items that the user interacted with. 
The following shows an example of the \textit{user history data structure}:
\begin{align*}
&{user_1: (time_1, item_1), (time_4, item_{10}), (time_{500}, item_{20})}\\
&{user_2: (time_3, item_{100}), (time_{20}, item_{100})}\\
\vdots\\
&{user_X: (time_5, item_{75}), (time_{20}, item_{50}),}\\
&{(time_{100}, item_{75}), (time_{400}, item_{1})(time_{420}, item_{10})}
\vspace{-4mm}
\end{align*}
Second, for each set of consecutive items purchased by any user, we create a list of users who have the same set of consecutive purchases in exactly that order. We refer to these sets of consecutive recent purchases as \textbf{\textit{keys}}. Multiple users may have the same key in their history. That is why each key keeps a \textit{list} of all the users that created the same key and the duration of the time they had the key. 
Here is an example of the \textit{recent item purchase history} when we consider the two most recent purchases. Each key consists of a pair of items. For instance, the first line shows item 1 and 10 were the most recent purchases of user 1 from time $4$ to time $500$.
\begin{align*}
&{key: \text{list of values}}\\
&{[item_1, item_{10}]: [user_1, time_4, time_{500}]}\\
&{[user_X, time_{420}, Current]}\\
&{[item_{10}, item_{20}]: [user_1, time_{1000}, Current]}\\
&{[item_{100}, item_{100}]: [user_2, time_{20}, Current]}\\
\vdots \\
& [item_{75}, item_{50}]:[user_X, time_{20}, time_{100}]\\
& [item_{50}, item_{75}]: [user_X, time_{100}, time_{400}]\\
& [item_{75}, item_{1}]: [user_X, time_{400}, time_{420}]
\end{align*}
The goal of the re-identification attack is to use only the $m$ ($m=2$ in the example above) most recent purchases by a user to track the user across different interaction sessions.
 To evaluate this attack:
\begin{enumerate}[noitemsep, leftmargin=*]
    \item We randomly select a timestamp and a user.
    \item For the selected user, we check the $m$ most recent purchases of the user at the selected timestamp and form a key = [recent purchase 1, recent purchase 2, ... recent purchase m] 
    \item We look up this key in the recent item purchase history dataset. If the same sequence of $m$ most recent items appear on another user at the same time window, this means these recent purchases are not unique for that specific user at that time and cannot be used as a fingerprint of a single user.
    \item On the other hand, if the $m$ item purchase history only belongs to that specific user, the duration of the time in which this key forms the most recent purchases of the user is extracted.
    \item We repeat this experiment for many random time stamps and random users to obtain $200,000$ samples.
\end{enumerate}
\begin{figure}[htbp]
 \centering
 \includegraphics[height=1.8in]{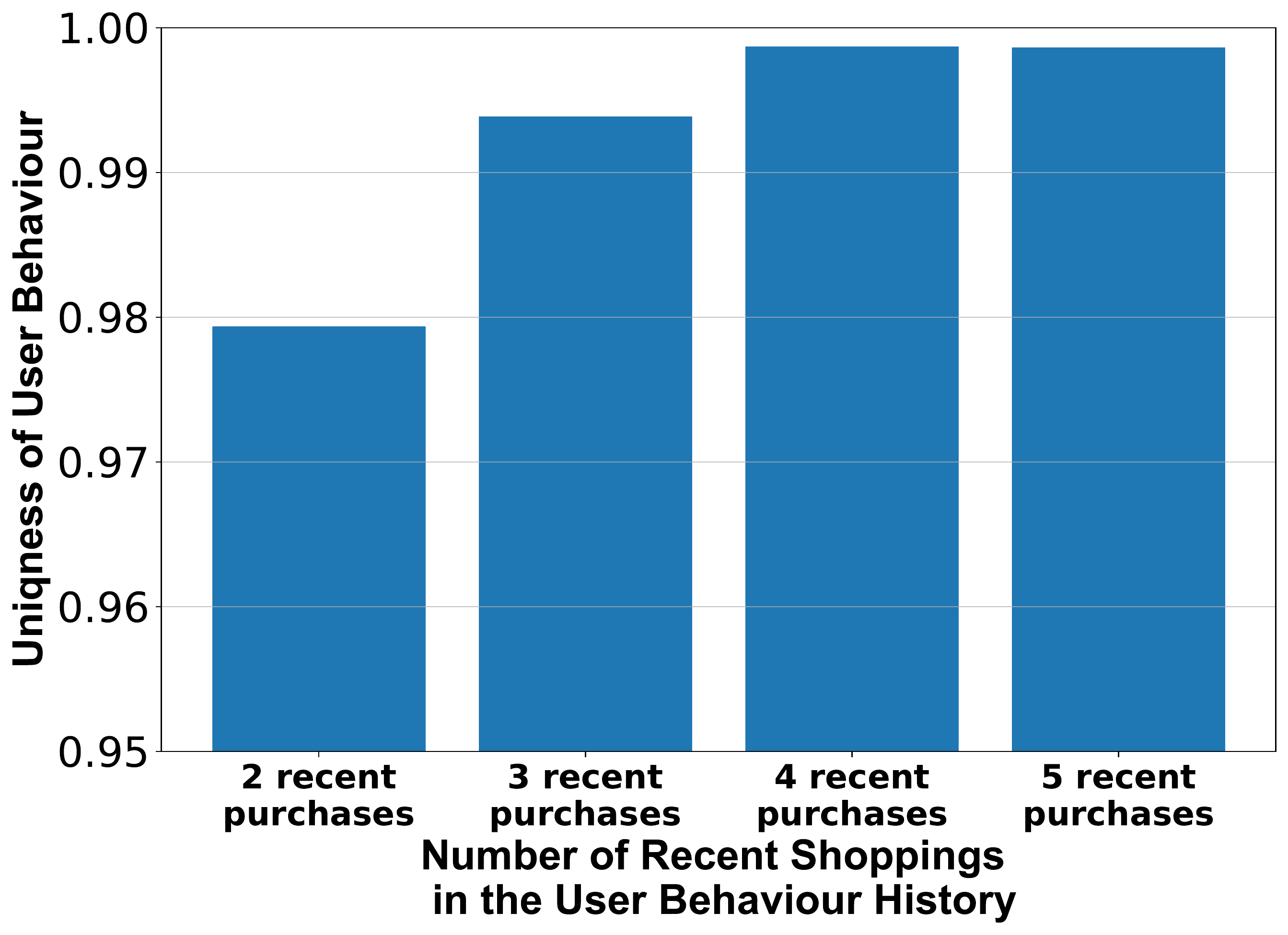}
  \vspace{-2mm}
\caption{Uniqueness of most recent purchases of users.}
 \label{fig:uniq}
\end{figure}
\begin{table}[tb]
\caption{Re-identification attack statistics about the number of keys and repeated keys.}
\label{tab:reident}
\centering
\resizebox{\columnwidth}{!}{%
\begin{tabular}{l ccccc}
\hline
{\begin{tabular}[c]{@{}l@{}}\bf {Number of} \\ \bf {recent purchases}\end{tabular}} & {\bf Number of users} &{\bf Number of keys} & {\begin{tabular}[c]{@{}l@{}} \bf {Total occurrences}\\ \bf{of keys}\end{tabular}} \\ \hline
$2$ &$898,803$ & $4,476,760$ & $8,114,860$ \\ 
$3$ &$799,475$ & $5,679,087$ & $7,216,057$ \\
$4$ &$705,888$& $5,587,578$ & $6,416,582$ \\
$5$ &$620,029$ & $5,197,043$ & $5,710,694$  \\
\hline
\centering
\end{tabular}
}
\end{table}
As depicted in Figure~\ref{fig:uniq}, we observe that even the two most recent purchases can serve as a unique identifier for $98\%$ of our samples. In other words, at a random point in time, the two most recent purchases of a user are unique for $98\%$ of randomly selected users. We performed a further analysis using more recent purchases and found that three, four, and five most recent purchases uniquely identify users with $99\%$ probability. 

\begin{figure}[t]
 \centering
\includegraphics[height=1.1in,width=1.6in]{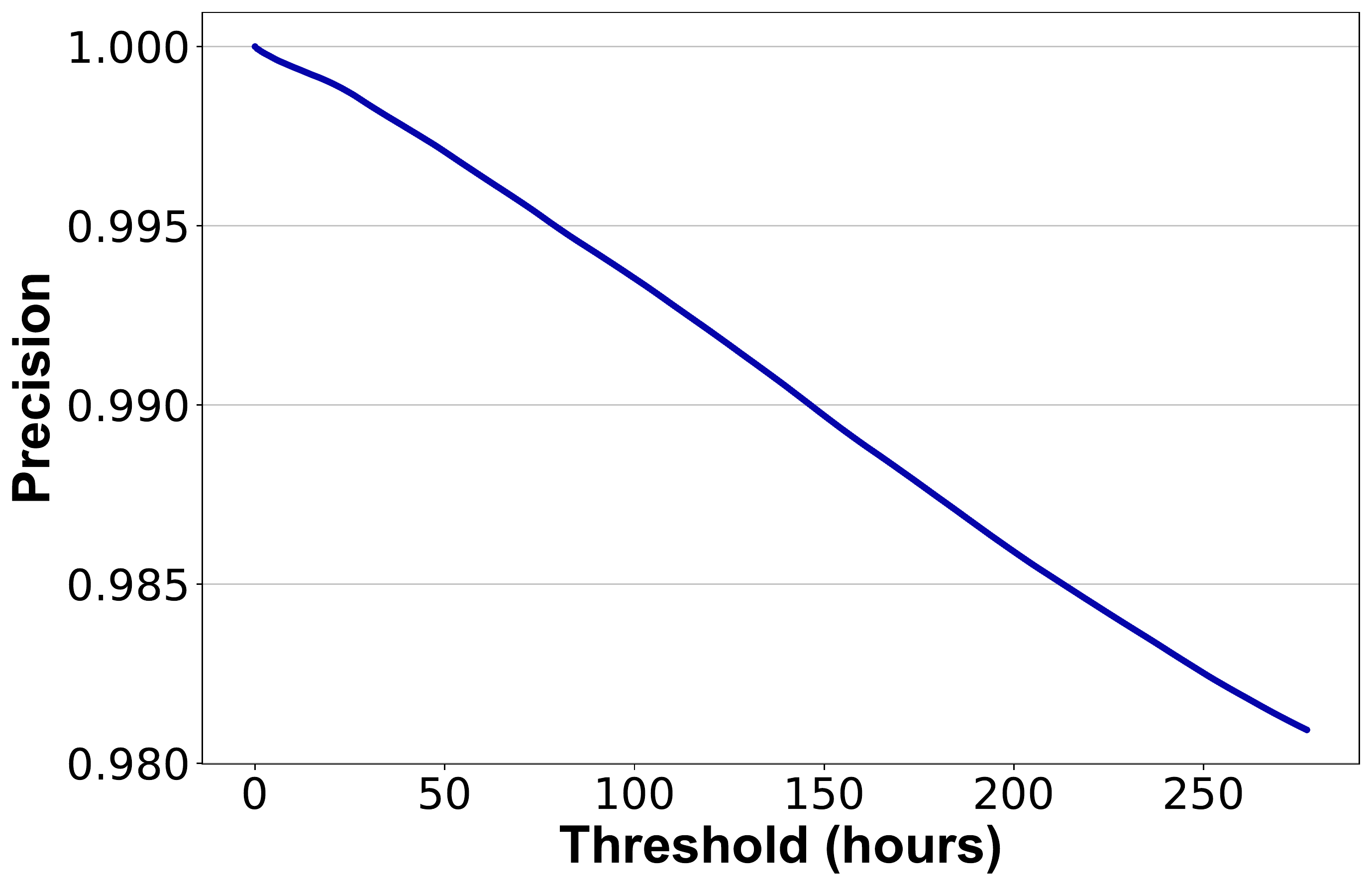}
\includegraphics[height=1.1in,width=1.6in]{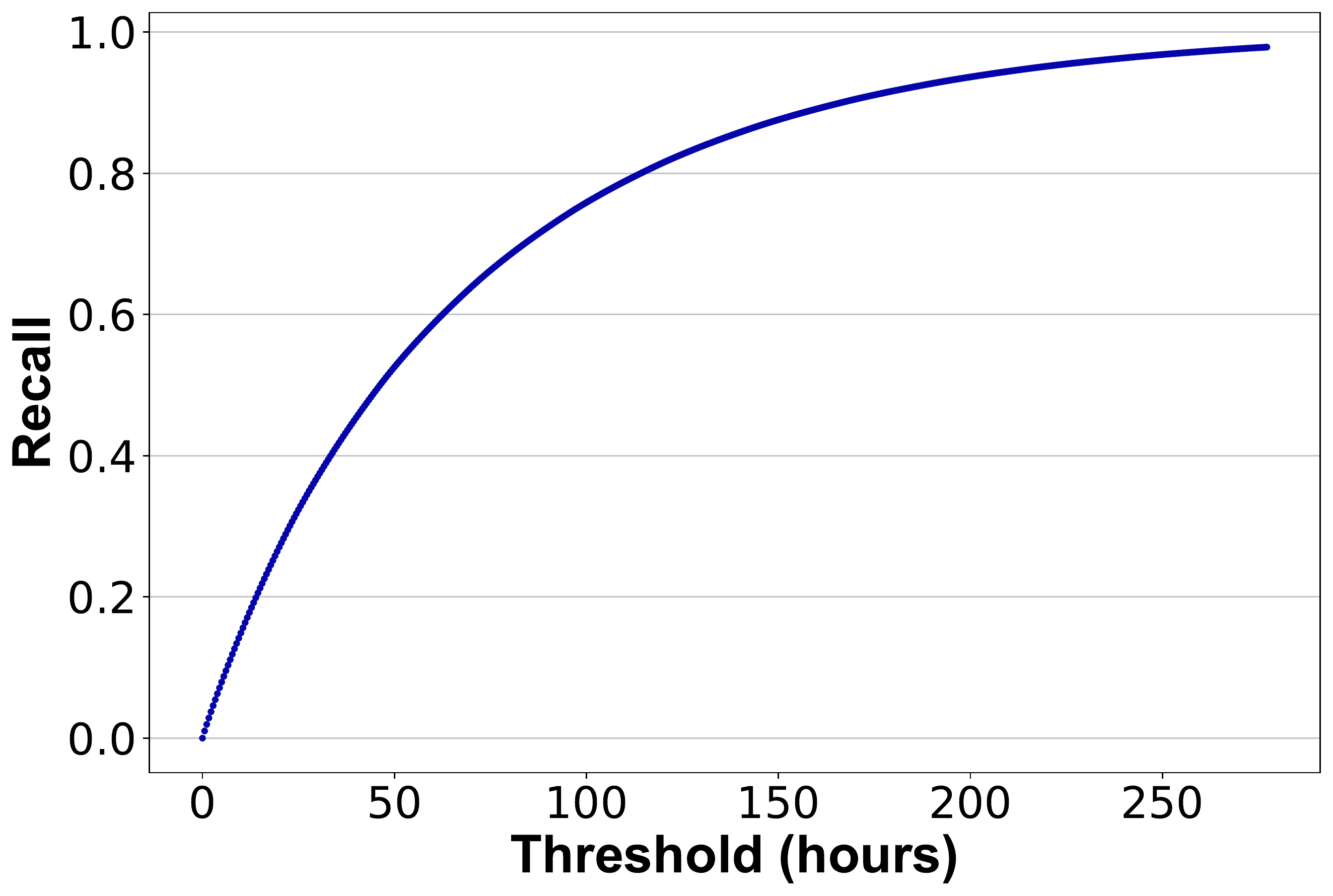}
\caption{Precision/recall trade-off based on different time threshold values.}
 \label{fig:recal}
\end{figure}

\noindent \textbf{Attack Method:} Up to this point, we demonstrated that recent purchases are highly unique for each user. We now discuss the attacks that aim to re-identify users by tracking their queries. Most recent items purchased by a user usually do not change with a very high frequency. For the period of time that these recent purchases remain the same, every query sent by the user has the same list of recent purchases. Therefore, the attacker is interested in using this knowledge to launch the attack. To accomplish this, the attacker first selects a time threshold. This time threshold is chosen to help the attacker to decide if the queries come from the same user or not. Meaning that if the time difference between receiving them is less than the time threshold and two distinct queries received by the cloud have the same most recent purchases, the attacker will predict that they come from the same user. Otherwise, it is assumed queries come from two different users.\\
\noindent \textbf{Evaluation Metric:} To measure the accuracy of this attack, we use the machine learning terms {\em precision} and {\em recall} defined in~\cite{buckland1994relationship} as shown in Eq~(\ref{eq:precision}). 
\begin{equation}
\label{eq:precision}
    Precision = \frac{TP}{(TP+FP)}, \quad Recall = \frac{TP}{(TP+FN)}~,
\end{equation}
where TP stands for \textbf{T}rue \textbf{P}ositives, FP represents \textbf{F}alse \textbf{P}ositives, and FN is \textbf{F}alse \textbf{N}egatives. Precision indicates what percentage of positive predictions are accurate and Recall indicates what percentage of actual positives are detected.\\
\noindent \textbf{Evaluation Result:} To evaluate the precision/recall tradeoff, we start from a very small time threshold and increase it gradually. As expected, with low time thresholds, precision is high with few false positives. But as the attacker increases the time threshold and can identify more of the actual positives (higher recall), the false positives increase as well, which reduces the precision. The reason for having more false positives with a large threshold is that, during a longer period of time, other users may generate the same key. Table~\ref{tab:reident} shows when the $2$ most recent purchases are used, there are around $4.5$ million keys but the total number of occurrences of these keys is around $8$ million times. This means for a fraction of the keys, the same keys are generated for different users at different times. These repeated keys are the source of false positives in our experiments. The decision of selecting the right threshold depends on the attacker's preference to have a higher recall or precision. Figure~\ref{fig:recal} shows this trade-off for different time threshold values. We gradually increase the time threshold from $1$ second to $277$ hours ($11.5$ days) and compute the recall and the precision. As shown in this figure, by increasing the time threshold to $11$ days recall will reach $1.0$ while there is an almost $0.02$ drop in precision. This means the attacker can link all the queries with the same recent purchases that come from the same users correctly. This comes at the cost of $2\%$ miss-prediction of the queries that do not come from the same user and only generates the same key at some point in their purchase history. \textit{These high precision and recall values, indicates how an attacker can track users who send queries to the recommendation model over time.}

\section{Data Pipeline in Production-Scale Recommendation Systems}
\label{sec:pipeline}

As mentioned earlier, exposing raw values of sparse features can leak sensitive information of a user. In this section, we discuss the current production-scale data pipeline for sparse feature processing and how such real system designs may impact the information leak.

One challenge in designing efficient embedding tables is that the values of sparse features may be unbounded, resulting in very large embedding table sizes. Consider the news articles produced in the world as a dynamic sparse feature item that a user may interact with. There are thousands of news articles in just a day from around the world and creating embeddings for each news item in an embedding table is impractically large.  For instance, the DLRM  recommendation model in 2021 needs 16x larger memory, compared to the one used in 2017~\cite{lui2021understanding, sethi2022recshard}. Furthermore, $99\%$ of model parameters belong to embedding tables~\cite{gupta2020architectural}. That is why production-scale models demand 10s of TB memory capacity~\cite{Mudigere2021, sethi2022recshard}. 
One common solution for converting high dimensional data to a low-level representation is to use hashing~\cite{shi2009hash}. Using hashing for recommendation systems was first suggested in~\cite{zhang2018efficient}. In addition to bounding sparse features to a fixed size, hashing helps with responding to the rare inputs that are not seen before~\cite{acun2021understanding, kang2020learning}. Furthermore, using high-cardinality features may cause over-fitting problems due to over parameterization~\cite{liu2020learnable, kang2020learning}. Considering all these reasons, sparse feature inputs in production-scale models are hashed prior to embedding look-ups.

In this section, we briefly explain how different hashing schemes work and then we analyze how hashing impacts information leakage. Recall that all the information leakage that we discussed in the prior sections is due to the fact that an adversary sees the raw value of embedding table indices. Now we analyze if embedding table hashing in recommendation systems, which is not necessarily designed for protecting data privacy, can help with reducing information leakage.

\subsection{Hash Functions}
There are multiple ways of reducing the embedding table size using hash functions, and they all have trade-offs. We explain some of the most common hashing schemes here.

\noindent\textbf{Embedding table as a hash-map}: With hash-map, embedding table entries are combined based on their similarity and a smaller embedding table is formed. However, to use the embedding table, a hash map should be kept to keep track of merged entries. This is the most accurate but the most expensive method in practice. In a previous study~\cite{zhang2018efficient}, the authors suggested that using locality sensitive hashing can approximately preserve similarities of data
while significantly reducing data dimensions. Frequency hashing~\cite{zhang2020model} also keeps a separate map with hot items and carefully maps only hot items to different entries in the table. This ensures that hot items do not collide, while items that are less frequently accessed may in fact be mapped to the same entry. \\
\textbf{Modulo hashing}: This is the cheapest and simplest hash to implement. This hashing performs modulo division based on the pre-defined size of the hash table. For hash size $P$, the hash function is as simple as $input \; mod \; P$. Though simple, it has the disadvantage that two completely different entities might collide.\\
\textbf{Cryptographic hashing}: This approach is a one-way cryptographic algorithm that maps an input of any size to a unique output of a fixed length of bits. A small change in the input drastically changes the output. Cryptographic hashing is a deterministic hashing mechanism.
\subsection{Statistical Analysis of Information Leakage After Hashing}

In this section, we analyze if the amount of randomization created by hashing can have any effect on reducing data leakage.
\begin{table*}[htb]
\caption{Entropy and mutual information analysis of pre-hash and post-hash embedding table indices.}
\label{tab:entropy}
\centering
\begin{tabular}{cc ccccc}
\hline
{\bf Dataset} & {\bf Table Name} & {\bf Original Table Size} & {\bf Post Hash Table Size}   & {\bf Pre-Hash Entropy} & {\bf Post-Hash Entropy} &{\bf MI}  \\ \hline
Taobao &Brands &$379,353$&$37,935$&$9.91$&$9.28$& $9.28$  \\ \hline
Taobao &Categories &$12,124$&$1,212$&$6.19$&$5.72$&$5.72$  \\
\hline
Kaggle &C3 &$1,761,917$ &$176,191$&$10.15$&$9.41$& $9.41$  \\ \hline
Kaggle &C18 &$4,836$&$483$&$5.92$&$5.27$& $5.27$  \\ \hline
Kaggle &C24 &$110,946$&$11,094$&$6.57$&$6.28$& $6.28$  \\ \hline
Criteo &C7 &$6,593$&$659$&$7.63$&$5.84$&$5.84$  \\
\hline
Criteo &C12 &$159,619$&$15,961$&$7.20$&$6.85$&$6.58$  \\
\hline
Criteo &C20 &$11,568,963$&$1,156,896$&$7.37$&$7.18$&$7.18$  \\
\hline
\centering
\end{tabular}
\end{table*}
In the following, we report our analysis of the entropy of pre-hash and post-hash indices as well as the mutual information analysis. Given a discrete random variable X, with possible outcomes: $x_1, \dots, x_n$  which occur with probability $p(x_1), \dots, p(x_n)$, the entropy is formally defined as~\cite{cover1999elements}:
\vspace{-2mm}
\begin{equation}
\label{eq:entorpy}
    H(X) = - \sum_{i=1}^N p(x_i)\times log(p(x_i))
\end{equation}
The binary (Base 2) logarithm gives the unit of bits (or "Shannons"). Entropy is often roughly used as a measure of unpredictability. In this part, we measure the entropy of the input and output of the hash function. In our specific evaluation, we first measure the probabilities in Eq~\eqref{eq:entorpy} by measuring the frequency of each outcome for pre-hash. We used modulo hash function for compressing the values and measured the post-hash frequencies. Finally, by applying Eq~\eqref{eq:entorpy}, we find out the amount of uncertainty in each of these values. As shown in Table~\ref{tab:entropy}, the pre-hash entropy of the brand table in Taobao dataset is almost 10 bits. Even after reducing the table size with hashing by 10 times, the amount of information is not reduced significantly for the post-hash values. For the category table, the amount of information was $6$ bits and it remains the same after 10 times reduction in the table size. For Kaggle, we selected three embedding tables with different sizes. C3 is the largest embedding table with $1,761,917$ entries. C18 represents the small tables with $4,836$ entries while C24 represents the moderate tables with $110,946$ entries. As shown in this table, the entropy of the sparse features varies between 10 bits to 6 bits depending on the feature. This entropy is not reduced significantly in the post hash values. Finally, the Criteo dataset is evaluated. Note that since the dataset is hashed differently, feature names are different from the Kaggle dataset. In this dataset, C7 is the smallest table with $6,593$ entries. C12 is the average-size table and C20 is the largest embedding table with $159,619$ and $11,568,963$ entries respectively. The details about embedding table sizes are reported in Appendix A. 
\textit{An important observation is that the entropy of information in indices is not reduced significantly after hashing. It implies that the post-hash indices hold almost the same amount of information as the pre-hash indices.}

\textbf{Mutual Information (MI) Analysis}
In probability and information theory, the mutual information of two random variables is a measure of the mutual dependence between the two variables. More specifically, it quantifies the "amount of information" obtained about one random variable by observing the other random variable. Mutual information between two random variables X and Y is measured by~\cite{cover1999elements}:
\begin{equation}
\label{eq:MI}
    I(X;Y) = H(X) - H(X|Y) = H(Y) - H(Y|X)
\end{equation}
Many prior works used MI as a measure of privacy guarantee~\cite{cuff2016differential, kalantari2017information, liao2017general, guo2020secure, mireshghallah2020not}.
In our example, we compute the mutual information between the pre-hash indices ($X$) and the post-hash indices ($Y$). Based on Eq(\ref{eq:MI}), the mutual information between post-hash and pre-hash indices is equal to the entropy of the post-hash indices (H(Y)) minus the conditional entropy of post-hash indices given the pre-hash indices $(H(Y|X))$. With deterministic hash functions, a post-hash index is deterministic for a given pre-hash index. This means there is no ambiguity in conditional entropy. So $H(Y|X)$ in Eq(~\ref{eq:entorpy}) is equal to zero and MI is equal to the entropy of post-hash indices. Our empirical result in Table~\ref{tab:entropy} also validates this point. \textit{Based on this observation, the mutual information between input and output of the hash is almost equal to the entropy of the hash input. This means that an adversary with unlimited computational power can recover almost all the information in the pre-hash indices by just observing the post-hash indices.}
\section{Frequency-based Attack}
\label{sec:design}
In the previous section, we demonstrated theoretically that the amount of information is not reduced after hashing. This section studies how an attacker can recover the raw values of sparse features even when hashing is used for embedding indices.
\begin{figure}[t]
 \centering
 \includegraphics[height=1.4in]{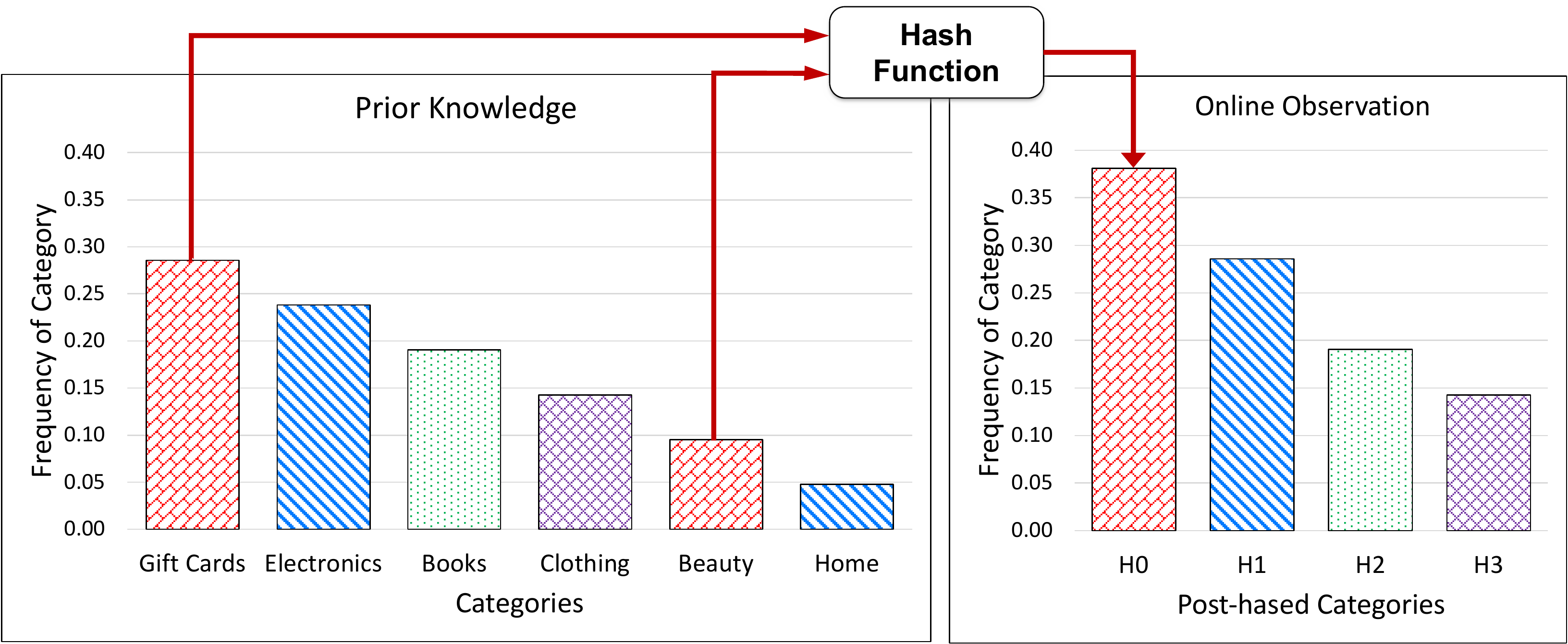}
 \caption{Frequency-based attack tries to reverse engineer the hash based on the frequency of accesses.}
 \label{fig:Hash}
\end{figure}
Through a hash function, users' raw data are remapped to post-hash values for indexing the embedding tables. An example of this is shown in Figure~\ref{fig:Hash}. \\
\noindent \textbf{Evaluation Setup:} In this section, we used three datasets to evaluate the attack. For Taobao dataset, we analyzed the two largest tables, category and brand. For Kaggle dataset, C3, C18, and C24 represent large, small, and moderate table sizes evaluated. Also for Criteo C7, C12, and C20 are evaluated for the same reason. For each dataset, we selected two disjoint random sets; training and test set. The training set samples form the prior distribution and the test sample are used for the evaluation as we explain in this section. Different configuration we used is reported in Table~\ref{tab:freq},~\ref{tab:freqkaggle}, and~\ref{tab:freqCriteo}.\\
\begin{table*}[htb]
\caption{Accuracy of hash inversion for the frequency-based attack for Taobao dataset.}
\label{tab:freq}
\centering
\resizebox{2\columnwidth}{!}{%
\begin{tabular}{l ccccccccccc}
\hline
{\begin{tabular}[c]{@{}l@{}} \bf Number of Samples used \\ \bf for Learning Distribution \end{tabular}} & {\begin{tabular}[c]{@{}l@{}} \bf Number of Samples\\ \bf for Evaluation \end{tabular}} & {\bf Top 1}   & {\bf Top 2} & {\bf Top 3} & {\bf Top 4} &  {\bf Top 5} & {\bf Top 6} & {\bf Top 7} & {\bf Top 8} & {\bf Top 9}& {\bf Top 10}  \\ \hline
{1,000,000}    & {1,000}   & $0.64$  & $0.76$ & $0.83$ & $0.87$ & $0.89$   &$0.90$ &$0.91$  & $0.92$ & $0.93$ & $0.94$  \\ 
{1,000,000}    & {100,000}   & $0.61$  & $0.75$ & $0.82$ & $0.86$ & $0.88$   &$0.90$ &$0.92$  & $0.92$ & $0.93$ & $0.93$  \\ 
{2,000,000}    & {100,000}   & $0.62$  & $0.76$ & $0.82$ & $0.86$ & $0.89$   &$0.91$ &$0.92$  & $0.93$ & $0.93$ & $0.94$   \\ 
{2,000,000}    & {1,000,000}   & $0.62$  & $0.76$ & $0.82$ & $0.86$ & $0.89$   &$0.91$ &$0.92$  & $0.93$ & $0.93$ & $0.94$   \\ 
\hline
\centering
\end{tabular}
}
\end{table*}
\noindent \textbf{Attack Method:} An adversary can launch attacks by collecting the frequency of observed indices, using prior knowledge about the distribution of feature values, and finding the mapping between the input and output of the hash. 
Here we show how an attacker can compromise a system with hashed input values where the hash function is $output = ( input+mask_{add})~ mod ~  P$ and $P$ is the hash size. 
We denote the frequency of possible input to a hash function by $x_1,x_2, \dots, x_N$ for N possible scenarios and its output frequency by $y_1, y_2, \dots, y_P$ of a hash size P.
We form the matrix $M \in \mathbb{R}^{P \times P}$ in which each column represents a different value for Mask ($[0,P-1]$). Basically, for each value of a mask, we compute the frequency of outcomes and form this Matrix. As shown, by increasing the value of the mask by 1, the column values are shifted. Hence, the Matrix M is a Toeplitz Matrix. Since a single column in this matrix is shifted and repeated the order of forming this matrix is $O(P)$.  
\begin{equation}
\mathbf{M} = 
\begin{bmatrix}
y_{1} & y_{P} & \cdots & y_{2} \\
y_{2} & y_{1} & \cdots & y_{3} \\
\vdots  & \vdots  & \ddots & \vdots  \\
y_{P} & y_{P-1} & \cdots & y_1 
\end{bmatrix}_{P \times P}
\end{equation}
The attacker's goal here is to invert the hash using the input distribution and its observation of the output distribution. Note an input dataset and an output dataset should be independent.
We define $\mathbf{a}_t$ as the distribution of embedding table accesses (post-hash) at time t. To reverse engineer the mask, an attacker has to find out which mask is used by the hash function. To do so, the attacker has to solve the optimization problem in Eq(~\ref{eq:min}).   
\begin{equation}
\label{eq:min}
\min_i \quad \norm{(\mathbf{m}_i - \mathbf{a}_t)}^2 = \min_i ( \norm{\mathbf{m}_i}^2 + \norm{\mathbf{a}_t}^2 - 2\mathbf{m}_i^\intercal   \mathbf{a}_t) 
\end{equation}
In Eq~\eqref{eq:min}, $\mathbf m_i$ represents the vector containing the frequencies of output values when mask $i$ is used. So its absolute value will be a constant one. This is similar for $\norm{\mathbf{a}_t}$. As a result, the optimization problem can be simplified to Eq(\ref{eq:opt}). 
\begin{align}
\label{eq:opt}
 &\bar P = \operatorname*{arg\,max}_i ( \mathbf{m_i}^\intercal \mathbf{a}_t ) \quad for\quad i\in [0,P-1] \implies \nonumber \\
  & \bar P = \operatorname*{arg\,max}_i (\mathbf{M}^\intercal\mathbf{a}_t) 
\end{align}
Usually, the order of computing such a matrix-vector product is $O(P^2)$. However, because $\mathbf M$ is a Toeplitz matrix, this matrix vector computation can be done in time complexity of $O(P \log{P})$~\cite{strang1986proposal}.
In order to implement this attack, we created two disjoint sets. The first set is used to extract the distribution (known distribution) and the second set is used for frequency matching and evaluating the frequency-based attack. In the first step, attackers try to reverse engineer the hash function and find the key based on the frequency matching. The attacker was able to successfully reverse engineer the hash and find the key based on the method described above. Note that the size of the output of the hash is smaller than the input size. This means multiple inputs map to the same output.

In the second step, the attacker tries to reverse engineer the post-hash indices and find out the value of raw sparse features. In order to do so, after finding the key of the hash, the attacker reverse engineer the post-hash value to the top most frequent pre-hash values based on the input distributions. 
\begin{table*}[htb]
\caption{Accuracy of hash inversion for the frequency-based attack for Kaggle dataset.}
\label{tab:freqkaggle}
\centering
\resizebox{2\columnwidth}{!}{%
\begin{tabular}{l l l cccccccccc}
\hline
{\begin{tabular}[c]{@{}l@{}} \bf Number of Samples used \\ \bf for Learning Distribution \end{tabular}} & {\begin{tabular}[c]{@{}l@{}} \bf Number of Samples\\ \bf for Evaluation \end{tabular}} & \textbf{Feature} & {\bf Top 1}   & {\bf Top 2} & {\bf Top 3} & {\bf Top 4} &  {\bf Top 5} & {\bf Top 6} & {\bf Top 7} & {\bf Top 8} & {\bf Top 9}& {\bf Top 10}  \\ \hline
{$100,000$}    & {$1,000$}  & C3 & $0.55$  & $0.55$ & $0.55$ & $0.55$ & $0.55$   &$0.55$ &$0.55$  & $0.55$ & $0.55$ & $0.55$  \\ 

{$100,000$}    & {$1,000$}  & C18 & $0.74$  & $0.90$ & $0.95$ & $0.96$ & $0.98$   &$0.98$ &$0.98$  & $0.98$ & $0.98$ & $0.98$  \\ 

{$100,000$}    & {$1,000$}  & C24 & $0.87$  & $0.92$ & $0.92$ & $0.92$ & $0.93$   &$0.93$ &$0.93$  & $0.93$ & $093$ & $0.93$  \\ 

{$1000,000$}    & {$10,000$}  & C3 & $0.63$  & $0.64$ & $0.65$ & $0.65$ & $0.65$   &$0.65$ &$0.65$  & $0.65$ & $0.65$ & $0.65$  \\ 

{$1000,000$}    & {$10,000$}  & C18 & $0.75$  & $0.89$ & $0.94$ & $0.96$ & $0.98$   &$0.98$ &$0.98$  & $0.99$ & $0.99$ & $0.99$  \\ 

{$1000,000$}    & {$10,000$}  & C24 & $0.90$  & $0.95$ & $0.96$ & $0.97$ & $0.97$   &$0.97$ &$0.97$  & $0.97$ & $097$ & $0.97$  \\ 

{$4,000,000$}    & {$100,000$}  & C3 & $0.68$  & $0.71$ & $0.71$ & $0.72$ & $0.72$   &$0.73$ &$0.73$  & $0.73$ & $0.74$ & $0.74$  \\

{$4,000,000$}    & {$100,000$}  & C18 & $0.78$  & $0.91$ & $0.95$ & $0.97$ & $0.98$   &$0.99$ &$0.99$  & $0.99$ & $0.99$ & $0.99$  \\ 

{$4,000,000$}    & {$100,000$}  & C24 & $0.91$  & $0.95$ & $0.97$ & $0.97$ & $0.98$   &$0.98$ &$0.98$  & $0.98$ & $0.98$ & $0.98$  \\ 

\hline
\centering
\end{tabular}
}
\end{table*}
\begin{table*}[htb]
\caption{Accuracy of hash inversion for the frequency-based attack for Criteo dataset.}
\label{tab:freqCriteo}
\centering
\resizebox{2\columnwidth}{!}{%
\begin{tabular}{l l l cccccccccc}
\hline
{\begin{tabular}[c]{@{}l@{}} \bf Number of Samples used \\ \bf for Learning Distribution\end{tabular}} & {\begin{tabular}[c]{@{}l@{}} \bf Number of Samples\\ \bf for Evaluation \end{tabular}} & \textbf{Feature} & {\bf Top 1}   & {\bf Top 2} & {\bf Top 3} & {\bf Top 4} &  {\bf Top 5} & {\bf Top 6} & {\bf Top 7} & {\bf Top 8} & {\bf Top 9}& {\bf Top 10}  \\ \hline
{$3,000,000$}    & {$200,000$}  & C7 & $0.33$  & $0.48$ & $0.61$ & $0.68$ & $0.74$   &$0.80$ &$0.84$  & $0.88$ & $0.91$ & $0.93$  \\

{$3,000,000$}    & {$200,000$}   & C12 & $0.89$  & $0.96$ & $0.98$ & $0.99$ & $0.99$   &$0.99$ &$0.99$  & $0.99$ & $0.99$ & $0.99$  \\ 

{$3,000,000$}    & {$200,000$}  & C20 & $0.93$  & $0.98$ & $0.99$ & $0.99$ & $1$   &$1$ &$1$  & $1$ & $1$ & $1$  \\ 

{$30,000,000$}    & {$2,000,000$}   & C7 & $0.33$  & $0.48$ & $0.58$ & $0.65$ & $0.73$   &$0.80$ &$0.85$  & $0.88$ & $0.92$ & $0.93$  \\ 

{$30,000,000$}    & {$2,000,000$}  & C12 & $0.89$  & $0.96$ & $0.98$ & $0.98$ & $0.99$   &$0.99$ &$0.99$  & $0.99$ & $0.99$ & $0.99$  \\ 

{$30,000,000$}    & {$2,000,000$}  & C20 & $0.85$  & $0.88$ & $0.91$ & $0.94$ & $0.96$   &$0.98$ &$0.99$  & $0.99$ & $0.99$ & $0.99$  \\

{$400,000,000$}    & {$4,000,000$} & C7 & $0.33$  & $0.48$ & $0.58$ & $0.65$ & $0.73$   &$0.80$ &$0.83$  & $0.88$ & $0.90$ & $0.93$  \\

{$400,000,000$}    & {$4,000,000$}  & C12 & $0.89$  & $0.96$ & $0.98$ & $0.98$ & $0.99$   &$0.99$ &$0.99$  & $0.99$ & $0.99$ & $0.99$  \\ 

{$400,000,000$}    & {$4,000,000$}  & C20 & $0.84$  & $0.88$ & $0.90$ & $0.92$ & $0.95$   &$0.97$ &$0.98$  & $0.99$ & $0.99$ & $0.99$  \\  

\hline
\centering
\end{tabular}
}
\end{table*}
\noindent \textbf{Evaluation Metric:} Accuracy in this case is the probability that the attacker correctly identifies an input raw value from the post-hash value. Let the function $g(y)$ be the attacker's estimate of the input, given the output query $y$, as follows,
\begin{align}
    g(y)=&\arg\max_x \text{Prob}(x) \qquad \text{s.t. }\quad \hat h(x)=y~,
\end{align}
where $\hat h(x)$ is the attackers' estimation of the hash function. Using this definition, accuracy is defined as
\begin{align}
    \label{eq:acc}
    \text{Accuracy} = \text{Prob}_{x\sim \mathcal P_X}\left( x= g(h(x)) \right)~,
\end{align}
where $h(x)$ is the true hash function, and the probability is over the distribution of the input query.

We also use the notation of \textit{top K accuracy} in this section. Essentially top $K$ accuracy is the probability of the input query being among the top guesses of the attacker. To formally define this, we first denote the set $\hat{\mathcal S}(y)$ as,
\begin{align}
    \hat{\mathcal S}(y) = \{x~|~ \hat h (x) = y\}~,
\end{align}
which is the set of all possible inputs, given an output query $y$, based on the attacker's estimation of the hash function. We now define the set $g_K(y)$ to be the top $k$ members of the set $\hat{\mathcal S}(y)$ with the largest probability,
\vspace{-1mm}
\begin{align}
    g_K(y) = \{x\in \hat{\mathcal S}(y)| \text{Prob}(x) \text{ is in the top }K\text{ probabilities.} \}
\end{align}
This means that $g_K(y)$ is the set of the top $K$ attacker's guesses, of the input query. Now we can use the function $g_k(y)$ to formally define the top $K$ accuracy,
\begin{align}
    \text{Accuracy}_{\text {top }K} = \text{Prob}_{x\sim \mathcal P_X}\left( x\in g_K(h(x)) \right)~,
\end{align}
where $h(x)$ is the true hash function, and the probability is over the distribution of the input query.\\
\noindent \textbf{Evaluation Result:} As shown in Table~\ref{tab:freq}, we change the number of interactions in these test sets to see the accuracy of hash-inversion and the attacker could achieve up to $0.94$ top 10 accuracy for the Taobao dataset. 

In Table~\ref{tab:freqkaggle}, we show the accuracy of this attack model for the Kaggle dataset. As demonstrated in this table for small embedding tables (represented by C18), even a small sample of a prior distribution and online queries observed by an attacker can lead to a high inversion accuracy while for large tables (represented by C3) more accurate distributions are needed.

The evaluation for the Criteo dataset is reported in Table~\ref{tab:freqCriteo}. In this dataset C7 is the smallest table, C20 is the average-size table and C12 is the largest embedding table (More details about embedding table sizes are reported in Appendix A.). Criteo dataset also validates the same observation as previous datasets.   
\textit{The key observation here is that, if an attacker observes and collects the frequency of queries, they can reconstruct the values of raw features with high accuracy by knowing the distributions of the pre-hash values and type of the hash function.}

\section{Is Private Hash a Solution?}
\label{sec:privateHash}
Note that hash functions are currently used for reducing the sizes of embedding tables rather than designed for privacy purposes. But if a private hash function is employed, can it guarantee zero information leakage? In other words, using any random mapping between inputs and outputs of the hash, and if an attacker does not know the hash, can they find the mapping just by observing the frequency of the accesses?
To answer this question, we first use a simple greedy attack to demonstrate the leakage of information. Then we use a more sophisticated machine learning based optimization exploiting sequences of access to show how an attacker can achieve a high hash inversion accuracy even when the hash function is unknown. 

\begin{figure}[t]
 \centering
 \includegraphics[height=1.88in,width=2.4in]{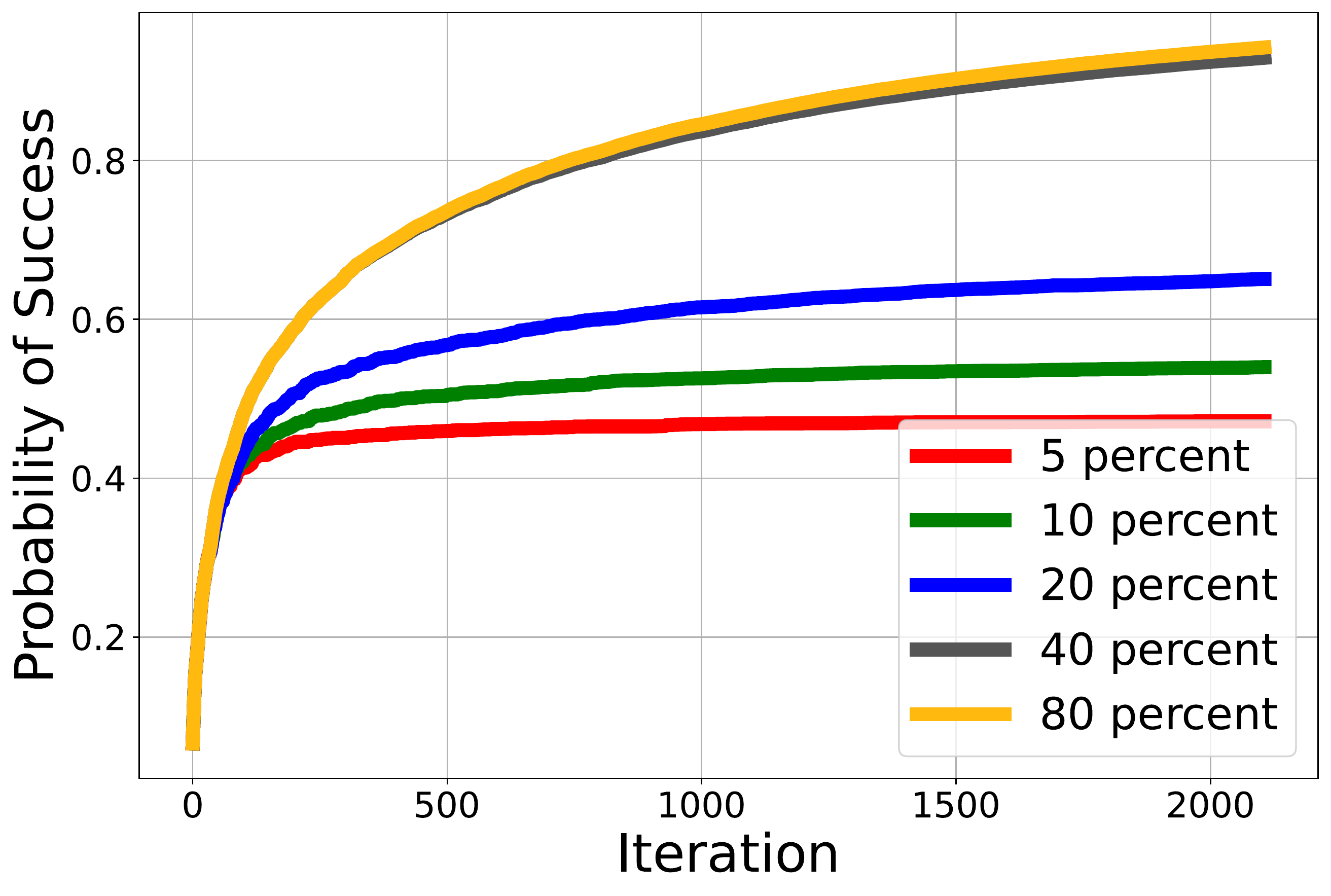}
 \vspace{-2mm}
 \caption{Hash-inversion accuracy increases with more optimization iterations and Larger hash sizes (Category Table). }
 \label{fig:private}
\end{figure}

\begin{figure}[t]
 \centering
 \includegraphics[height=1.88in,width=2.4in]{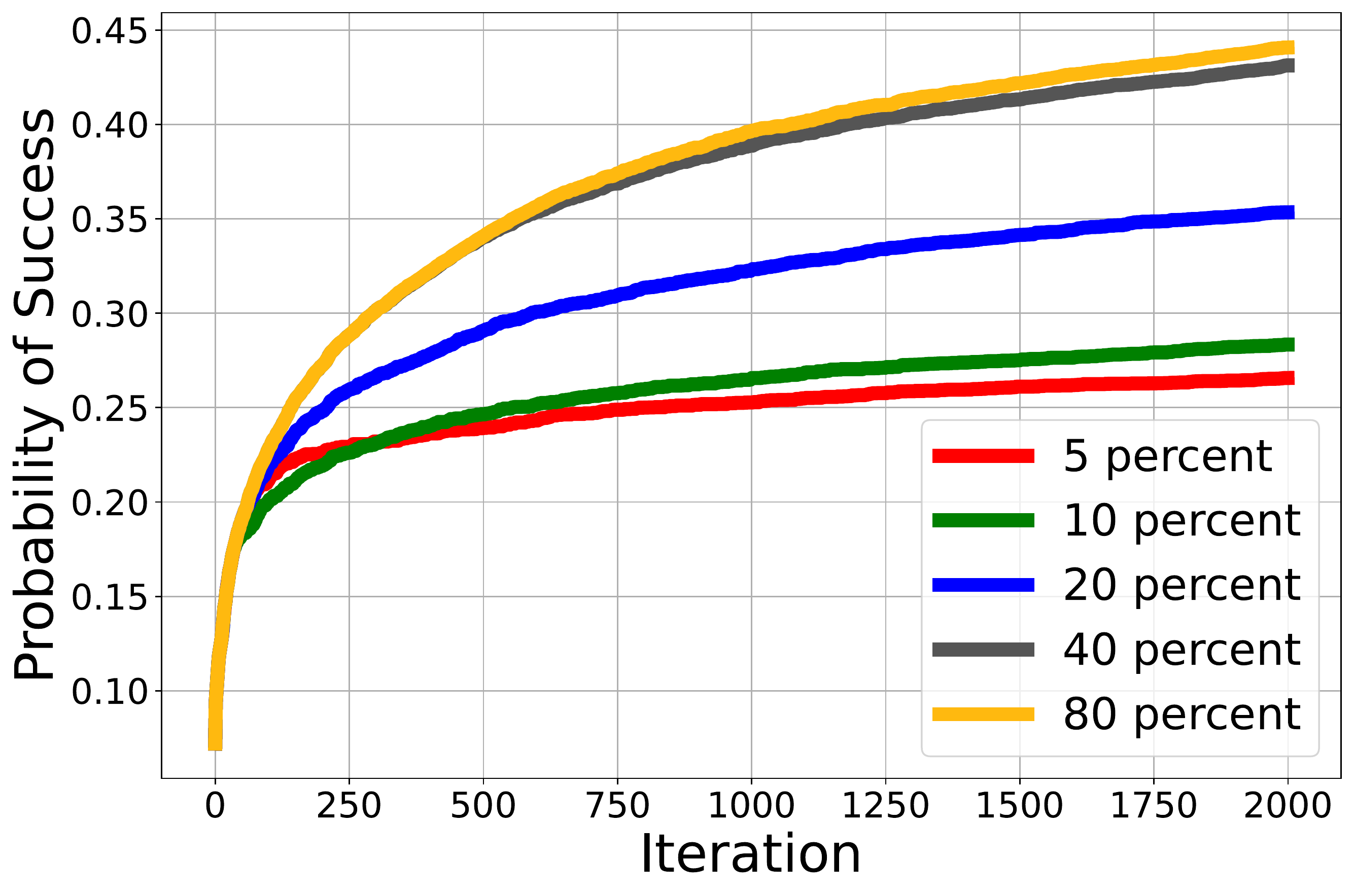}
  \vspace{-2mm}
 \caption{Hash-inversion accuracy increases with more optimization iterations and Larger hash sizes (Brand Table).}
 \label{fig:brandprivate}
\end{figure}

We first design a greedy attack to map the inputs and outputs by matching the frequencies without having any further information about the hash function. 
The only knowledge the attacker has are the prior distribution of pre-hash accesses and the observed post-hash access to the embedding table. We analyzed the category table of $12,000+$ pre-hash entries and $1,200$ post-hash entries ($P=0.1N$). We randomly map each of the $12,000$ inputs to an output. Then we launched the frequency-based attack without providing any information about this mapping to the attacker. This simple attack could successfully figure out the correct mapping for $23\%$ of the accesses. This analysis showed that although a private hash can reduce the amount of information leakage, it will not eliminate the leakage completely and is still susceptible to this type of attack. Now we take a step further to show how this attack can achieve an even higher inversion accuracy. \\
\noindent \textbf{Evaluation Setup:} As we explained in the previous sections, the user shares their most recent behaviors with the recommendation system to receive accurate suggestions. In this section, we show that the combination of the users' past shopping behaviors within one query, can help attackers launch more sophisticated attacks. Hence, for evaluating this attack we use Taobao dataset that provides this shopping behaviours. We evaluated both Category and Brand tables with more than $379$K and $12$K raw entries respectively.\\
\noindent \textbf{Attack Method:} Assume that $N$ is the size of the input, and $P$ is the size of the output of the hash, and the hash function $\mathbf h (.)$ maps the input to the output. Thus, $\mathbf h[i] = j$ means that the hash function, maps input index $i$ to output index $j$. We do not impose any assumptions on the hash function in this part. Assume that the joint distribution of the two most recent purchases of the input and the output are shown by the matrices $\mathbf X\in\mathbb R^{N\times N}$ and $\mathbf Y\in\mathbb R^{P\times P}$, respectively. This means that the probability of $(i_1,i_2)$ in the input is $\mathbf X_{i1,i2}$ and the probability of $(j_1,j_2)$ in the output is $\mathbf Y_{j1,j2}$. Also assume that the matrix $\mathbf B\in\mathbf R^{P\times N}$ is the one-hot representation of the hash function $\mathbf h(.)$, such that 
\vspace{-4mm}
\begin{align}
    \mathbf B_{j, i}=\bigg\{\begin{matrix}
    1\quad \mathbf h(i) = j\\
    0\quad\text{otherwise}
    \end{matrix}
\end{align}
Using these notations, we can show that,
\begin{align}\label{eq:initial_Form}
    \mathbf Y = \mathbf B\mathbf X\mathbf B^T~.
\end{align}
To prove this, note that
\begin{align}\label{eq:prove8}
   &\mathbf Y_{i_1, i_2} = \sum_{j_1,j_2}\mathbbm{1}_{\mathbf h(j_1)=i_1}\mathbbm{1}_{\mathbf h(j_2)=i_2}\mathbf X_{j_1,j_2} \nonumber \\
   & = \sum_{j_1, j_2} \mathbf B_{i_1,j_1}\mathbf X_{j_1,j_2}\mathbf B_{j_2,i_2}~,
\end{align}
where $\mathbbm 1_{\mathcal E}$ is the indicator function of the event $\mathcal E$, therefore $\mathbbm{1}_{\mathbf h(j_1)=i_1} = \mathbf B_{i_1, j_1}$. Eq~\eqref{eq:prove8} yields \eqref{eq:initial_Form}. Now, to estimate $\mathbf B$, we would like to ideally solve the following optimization.
\begin{align}\label{eq:secondary}
    \hat{\mathbf B} = \arg\min_{\mathbf B\in\mathcal B} \|\mathbf Y - \mathbf B\mathbf X\mathbf B^T\|^2_F~,
\end{align}
where $\|\mathbf X\|_F^2=\sum_{i,j}\mathbf X_{i,j}^2$ is the Frobenius norm and $\mathcal B$ is the space of all possible matrices $\mathbf B$, that represents a hash function. Optimization \eqref{eq:secondary} is an integer programming and NP-hard problem, due to the constraint in the minimization. To approximately solve this, we use Orthogonal Matching Pursuit (OMP)~\cite{tropp2007signal}. The idea behind OMP is to find one column of the matrix $\mathbf B$ in each iteration, in such a way that the new column satisfies the constraint on $\mathbf B$, and the new added column minimizes the loss function in \eqref{eq:secondary} the most (compared to any other feasible column). Note that in each iteration of our algorithm, we make sure that the matrix $\mathbf B$ can represent a hash function. The size of Matrix $\mathbf B$ can grow large based on the embedding table size. Thus, in our implementation we used CSR format since this matrix is sparse.\\ 
\noindent \textbf{Evaluation Metric:} Accuracy  is the probability that the attacker correctly identifies a raw input value from the post-hash value. We used top-1 accuracy which is defined in Eq~\eqref{eq:acc}.\\
\noindent \textbf{Evaluation Result:} To evaluate this attack, we measure the accuracy of the hash inversion function when changing the hash size. Figure~\ref{fig:private} demonstrates the hash-inversion accuracy using this optimization for the Taobao category table. We used different hash sizes to evaluate this attack. The size of the hash table changes from $0.05$  $(P=0.05N)$ of the original table size to $0.80$ of the table size. It shows how this accuracy increases over iterations until it saturates. For the large hash sizes, $P=0.8N$, accuracy reaches $94\%$, which means the this attack can recover raw values from hashed values for $94\%$ of accesses. Since the embedding table size for the Brand table is large, we used the Compressed Sparse Row (CSR) implementation to optimize the memory usage of the attacker. This way we could analyze the same attack on the brand embedding table with $379,353$ raw entries. Figure~\ref{fig:brandprivate} shows how different hash sizes can change the attacker's accuracy for hash inversion in the brand table. \textit{The key takeaway is that, even an unknown private hash cannot reduce the information leakage. An attacker can use this frequency-based machine learning optimization to recover the raw value features with high accuracy.}

\section{Implications for Private Recommendation Systems}
\label{sec:imp}

\begin{table*}[t]
\caption{Attack summary.}
\label{tab:summary}
\centering
\begin{tabular}{l lll}
\hline
{\bf Attack} & {\bf Goal} & {\bf Assumption} & {\bf Evaluation Metric}   \\ \hline
Identification & Finding the identity of users & \begin{tabular}[c]{@{}l@{}} Attacker observes accesses \\  Has prior knowledge about distribution of accesses \end{tabular}  & K-anonymity \\ \hline

Sensitive Attribute & Extracting sensitive user features& \begin{tabular}[c]{@{}l@{}} Attacker observes accesses \\ Has prior knowledge about distribution of accesses \end{tabular} & Ambiguity \\
\hline

Re-Identification & Tracking users over time & Attacker observes accesses & Precision and Recall \\
\hline

Frequency-based attack & Finding users' raw feature values& \begin{tabular}[c]{@{}l@{}} Attacker observes accesses \\ Has prior knowledge about distribution of accesses \\Knows hash function \\
Does not know secret key for has \end{tabular} & Inversion Accuracy \\
\hline
\begin{tabular}[c]{@{}l@{}} OMP-based frequency attack \\  for private hash\end{tabular}   & Finding users' raw feature values & \begin{tabular}[c]{@{}l@{}} Attacker observes accesses \\ Has prior knowledge about distribution of accesses \\No information about hash \end{tabular} & Inversion Accuracy \\
\hline
\centering
\end{tabular}
  \vspace{-6mm}
\end{table*}

Our threat model is based on the common practices employed by the industry's recommendation systems. They are typically deployed in the cloud for inference serving~\cite{niu2020billion}. 
In such a setting, a pre-trained model is hosted by a cloud server. The interaction history of each end user is kept in a user's local web browser or on a merchant's site where the merchant is precluded from sharing these data with other platforms without users' consent. This assumption is particularly important as it reflects the growing awareness in protecting personal data privacy. 

There are various techniques that protect computations on cloud systems. These techniques include fully homomorphic encryption (FHE)~\cite{shmueli2017secure}, multi-party computation (MPC)~\cite{goldreich1998secure}, and 
trusted execution environments (TEEs)~\cite{costan2016intel, ARMREALM}. 
However, none of these techniques protect the privacy of memory access patterns. 
For example, while Intel SGX protects computational confidentiality and integrity, it has been shown to be vulnerable to side-channel attacks via memory access pattern leakage~\cite{wang2017leaky}. 
This paper shows that the information leakage through embedding table accesses may be used to extract private user information, suggesting that memory access patterns need to be protected if strong privacy protection is necessary for recommendation systems in the cloud. 

Table~\ref{tab:summary} summarizes the attacks introduced in this paper. Each of them has a different goal. In all of these attacks, an attacker launches the attack by exploiting and analyzing the access patterns they observe. In some of the attacks, an attacker uses prior knowledge gleaned from the distribution of the accesses. In this work, we also define different metrics to evaluate each of these attacks. The high success rate of these attacks, highlights the importance of access pattern protection in the cloud-based recommendation systems.

\section{Related Work}
\label{sec:future}
The risk of information leakage in recommendation systems has been explored in prior works.  However, most of the research in this area focused on other models (e.g. content filtering) or dense features. Access pattern privacy in recommendation systems is a new topic and current Federated learning and Oblivious RAM schemes have shortcomings when it comes to DNN-based recommendation systems as we discuss here.

The study in~\cite{ zhang2021membership} designed a membership inference attack against a recommendation system to infer the training data in a content filtering model. Abdelberi {\em et al.} used a statistical learning model to find a connection between users' interests and the demographic information that users are not willing to share~\cite{chaabane2012you}.
Previous studies also investigated the risk of cross-system information exposure~\cite{chaum1985security, sweeney2002k}. For instance, a former Massachusetts Governor was identified in voter registration records by the combination of a zip code, birth date, and gender. Using this information, the researchers were able to identify him in a supposedly anonymous medical record dataset~\cite{sweeney2002k}. Most of the prior research in this domain was focused on information leakage through dense features~\cite{akhtar2018threat, choquette2021label, li2021membership, calandrino2011you, beigi2020survey}. 
Also, there are prior works investigating sparse feature leakage in other domains~\cite{ghinita2008anonymization, aggarwal2007privacy}. However, these leakages are through sparse feature values and not the embedding table accesses. Sparse feature's information leakage through embedding table accesses was explored for NLP models~\cite{song2020information, aggarwal2007privacy}. This attack aimed to disclose the embedding tables' input values based on their output which is different from our threat model. Access pattern attacks are also investigated in databases research ~\cite{grubbs2019learning, bindschaedler2017tao}. However, these attacks and defense schemes are fundamentally different from the ones in recommendation systems. In database attacks, the goal is to find the value of the encrypted data of the database based on the range queries or the correlation of different rows.

Using federated learning for training centralized recommendation models has gained attention recently~\cite{yao2021device, yang2020federated}. One of the problems of using federated learning for recommendation systems is the large size of embedding tables. These schemes usually use decomposition techniques such as tensor train to fit embedding tables on the edge devices~\cite{oseledets2011tensor}. However, because of the accuracy drop, the compression ratio is not high which makes them incompatible with edge devices. TT-Rec mitigates the performance degradation of tensor decomposition by initializing weight tensors by Gaussian distribution~\cite{yin2021tt}. Niu {\em et al.} proposed an FL framework to perform a secure federated sub-model training~\cite{niu2020billion}. They employed Bloom filter, secure aggregation, and randomized response to protect users' private information. But, inference solutions are not discussed in these federated learning approaches. DeepRec~\cite{han2021deeprec} proposed an on-device recommendation model for RNNs. In this work, there is a global model trained by public data that is available from before GDPR. Each device downloads this global model and re-train the last layer with their data. 
The problem with this model is that it depends on before GDPR public data. However, with new models come new features, which were not collected before. Thus they can not rely on this scheme for future models. 

One approach to obfuscating the embedded table access pattern is to use Oblivious RAM (ORAM)~\cite{goldreich1996software, stefanov2018path, ren2014ring}. At a high level, for each read or write operation, ORAM controller reads and writes not only the requested block, but also many random blocks. In this way, ORAM hides the information about real blocks from the attacker.  However, the overhead of ORAM is unlikely to be acceptable for real-time applications such as recommendation system inference due to Service Level Agreement (SLA)~\cite{hazelwood2018applied}. Even the most optimized version of ORAM suffers from 8-10 times performance overhead~\cite{raoufi2022ir}. A previous study~\cite{rajat2021look} tries to optimize ORAM for recommendation systems training. But, the scheme relied on pre-determined sequence of accesses in training and is not applicable to inference. In our future work, we plan to investigate low-latency protection schemes for embedding table accesses.

\vspace{-1mm}
\section{Conclusion}
\vspace{-1mm}
\label{sec:con}
In this work, we shed light on the information leakage through sparse features in deep learning-based recommendation systems. 
Our work pivoted the prior investigation focus on dense feature protection to the unprotected access patterns of sparse features. 
Some of these attacks such as identification and sensitive attribute attacks were investigated in other models, but to the best of our knowledge, this is the first effort analyzing these vulnerabilities in deep learning-based recommendation models. Furthermore, re-identification and frequency-based attacks are only explored in other domains. This paper pivots these attacks to the recommendation system domain. Also, we show that even secret hash functions cannot solve the privacy issue.
\cleardoublepage
\newpage

\bibliographystyle{IEEEtran}
\bibliography{sample-base}
\begin{table*}[ht]
\vspace{-170mm}
\caption{Embedding table information in Kaggle dataset.}
\label{tab:kaggle}
\centering
\resizebox{2\columnwidth}{!}{%
\begin{tabular}{lcccccccccccccccccccccccccc}
\hline
\textbf{Features} & \textbf{C1} & \textbf{C2} & \textbf{C3} & \textbf{C4} &
\textbf{C5} & \textbf{C6} & \textbf{C7} & \textbf{C8} &
\textbf{C9} & \textbf{C10} & \textbf{C11} & \textbf{C12} &
\textbf{C13} & \textbf{C14} & \textbf{C15} & \textbf{C16} &
\textbf{C17} & \textbf{C18} & \textbf{C19} & \textbf{C20} &
\textbf{C21} & \textbf{C22} & \textbf{C23} & \textbf{C24} & \textbf{C25} & \textbf{C26} \\
\hline

\textbf{Size}&$1396$& $550$&$1761917$&$507795$&$290$&$21$&$11948$ &     $608$&$3$&$58176$&$5237$&$1497287$&$3127$&$26$&$12153$&$1068715$ &      $10$&$4836$&$2085$&$4$&$1312273$&$17$&$15$&$110946$&$91$&$72656$ \\ 
\hline
\end{tabular}
}
\vspace{-370mm}
\centering
\end{table*}
\vspace{-60mm}
\begin{table*}[ht]
\caption{Embedding table information in Criteo dataset.}
\label{tab:Criteo}
\centering
\resizebox{2\columnwidth}{!}{%
\begin{tabular}{lcccccccccccccccccccccccccc}
\hline
\textbf{Features}&\textbf{C1} & \textbf{C2} & \textbf{C3} & \textbf{C4} &
\textbf{C5} & \textbf{C6} & \textbf{C7} & \textbf{C8} &
\textbf{C9} & \textbf{C10} & \textbf{C11} & \textbf{C12} &
\textbf{C13} & \textbf{C14} & \textbf{C15} & \textbf{C16} &
\textbf{C17} & \textbf{C18} & \textbf{C19} & \textbf{C20} &
\textbf{C21} & \textbf{C22} & \textbf{C23} & \textbf{C24} & \textbf{C25} & \textbf{C26} \\
\hline

\textbf{Size}&$10663931$ & $27772$ & $15365$ & $7184$ & $19519$ & $3$ & $6593$ & $1313$ & $63$ & $6690695$ & $469071$ & $159619$ & $10$ & $2207$ & $9382$ & $72$ & $4$ & $953$ & $14$ & $11568963$ & $2536927$ & $9187074$ & $230664$ & $10391$ & $94$ & $94$ \\ 
\hline
\end{tabular}
}
\centering
\end{table*}

\end{document}